# STELLAR MODEL ANALYSIS OF THE OSCILLATION SPECTRUM OF η BOOTIS OBTAINED FROM MOST

D. B. GUENTHER

Institute for Computational Astrophysics, Department of Astronomy and Physics, Saint Mary's University, Halifax, NS, Canada, B3H 3C3

T. KALLINGER, P. REEGEN, W. W. WEISS

Institut für Astronomie, Universität Wien Türkenschanzstrasse 17, A–1180 Wien, Austria

J. M. MATTHEWS, R. KUSCHNIG

Dept. Physics and Astronomy, Univ. of British Columbia, 6224 Agricultural Road, Vancouver, BC, Canada, B6T 1Z1

S. MARCHENKO

Dept. of Physics and Astronomy, Western Kentucky University, 1 Big Red Way, Bowling Green, Kentucky 42101 USA

A. F. J. MOFFAT

Obs du mont Mégantic, Dépt de physique, Univ de Montéal C. P. 6128, Succ. Centre-Ville, Montréal, QC, Canada, H3C 3J7

S. M. RUCINSKI

Dept. Astronomy & Physics, David Dunlap Obs., Univ. Toronto, P. O. Box 360, Richmond Hill, ON, Canada, L4C 4Y6

D. SASSELOV

Harvard-Smithsonian Center for Astrophysics, 60 Garden Street, Cambridge, MA 02138, USA

G. A. H. WALKER

Dept. Physics and Astronomy, Univ. of British Columbia, 6224 Agricultural Road, Vancouver, BC, Canada, B6T 1Z1

ABSTRACT

Eight consecutive low-frequency radial *p*-modes are identified in the G0 IV star η Bootis based on 27 days of ultraprecise rapid photometry obtained by the MOST (Microvariability & Oscillations of Stars) satellite. The MOST data extend smoothly to lower overtones the sequence of radial *p*-modes reported in earlier groundbased spectroscopy by other groups. The sampling is nearly continuous, hence, the ambiguities in *p*-mode identifications due to aliases, such as the cycle/day alias found in ground observations, are not an issue. The lower-overtone modes from the MOST data constrain the interior structure of the model of η Boo, giving a best fit on a grid of ~300,000 stellar models for a composition of $(X, Z) = (0.71, 0.04)$, a mass of $M = 1.71 \pm 0.05$ M$_\odot$, and an age of $t = 2.40 \pm 0.03$ Gyr. The surface temperature and luminosity of this model, which was constrained only by using the oscillation modes, is close (one sigma) to current best estimates of η Boo's surface temperature and luminosity. With the interior fit anchored by the lower-overtone modes seen by MOST, standard models are not able to fit the higher-overtone modes with the same level of accuracy. The discrepancy, model minus observed frequency, increases from 0.5 *μ*Hz at 250 *μ*Hz to 5 *μ*Hz at 1000 *μ*Hz, and is similar to the discrepancy that exists between the Sun's observed *p*-mode frequencies and the *p*-mode frequencies of the standard solar model. This discrepancy promises to be a powerful constraint on models of 3D convection.

*Subject headings*: stars: individual (η Bootis) — stars: interiors — stars: oscillations

## 1. INTRODUCTION

Rigorous asteroseismic tests of the physics of stellar evolution in stars like η Boo require unambiguous mode identifications and accurate frequencies (±1 *μ*Hz or less), and are aided by the detection of lower-overtone modes. From the ground, although it is possible to obtain the long observing runs required to resolve closely spaced frequencies, the runs are typically broken by the day-night cycle which introduces 1 cycle/day (11.57 uHz) aliasing artifacts into the power spectrum of the time series. This leads to ambiguities in mode identification, especially when trying to identify nonradial modes which are subject to mode bumping.

Evidence for the existence of *p*-modes on η Boo, based on groundbased spectroscopic observations, has been reported by Kjeldsen et al. (1995), Kjeldsen et al. (2003, hereafter K2003), and Carrier et al. (2005, hereafter C2005). The K2003 mode identifications are based on four separate data sets: 1. new equivalent width measurements by the authors, 2. the Kjeldsen et al. (1995) equivalent width measurements, 3. new Doppler shift measurements by the authors, and 4. Doppler shift measurements by Brown et al. (1977). The C2005 mode identifications are based on Doppler shift measurements made at two sites operating simultaneously. While the existing groundbased oscillation data on η Boo are impressive, they do not yet have sufficient completeness or sensitivity to lower-frequency variations to permit critical testing of the models. Only a few of the frequencies of the 21 peaks identified as modes in K2003 coincide with the frequencies of the 21 peaks identified as modes in C2005 (see Fig. 6 C2005), within the ~1.5 *μ*Hz uncertainty of the observations. Also, because of the daily gaps in the time series 1 cycle/day alias corrections had to be applied arbitrarily in K2003 (to 6 frequencies out of the 8 identified as radial modes) and in C2005 (to 4 out of 8) to produce plausible *p*-mode echelle diagrams with their observed power spectra.

From a modeling point of view the bright G0 IV star η Bootis is especially interesting. Stellar models fit to η Boo's observed position in the HR-diagram show that η Boo's convective envelope is very thin, containing less than 1% of the total mass of the star. This could allow *g*-modes, driven in the interior, to tunnel through the evanescent convective envelope to the surface where they may be observable. In addition, the distinct physical characteristics of the thin convective envelope provide a new datum to the study of turbulent convection and its driving of *p*-mode oscillations.

We also know from our models of η Boo's oscillation spectrum that its nonradial *p*-modes should be mixed, having *g*-mode-like character in the interior and *p*-mode-like character near the surface (Christensen-Dalsgaard et al. 1995; Guenther & Demarque 1996; Guenther 2004; Guenther 2005).

Mixed modes occur at frequencies where the *g*-mode eigenspectrum encroaches on the *p*-mode eigenspectrum. When the frequency of a *g*-mode nears the frequency of a *p*-mode (of similar degree *l*), the modes couple and the resultant frequency of the mixed mode is perturbed from the *p*-mode's regular near-equidistant spacing from its neighbours. In some cases, this frequency shift can be as large as the spacing itself. This effect is also referred to as mode bumping and avoided crossings.

In this paper we describe our preliminary analysis of the observations of η Boo made with the MOST[1] (Microvariability & Oscillations of STars) microsatellite. MOST (Walker et al. 2004) houses a 15-cm telescope with an optical bandpass (350 - 750 nm) feeding a stable CCD photometer. Its Sun-synchronous 820-km polar orbit enables MOST to stare at certain stars for up to 8 weeks without interruption. The orbital period is 101.413 minutes. The combination of instrument design and orbit makes MOST theoretically capable of detecting photometric oscillations with amplitudes of a few parts per million (ppm) across a wide frequency range. The noise level, which varies with target brightness and the viewing angle of the satellite relative to the Sun and illuminated Earth ultimately sets the lower detection limit, which in the case of η Boo is ~10 ppm.

The scientific results obtained by MOST since its launch on 30 June 2003 include: the null detection of *p*-modes on Procyon (Matthews et al. 2004); measurement of differential rotation via starspots on $\kappa^1$ Ceti (Rucinski et al. 2004); and the interaction of rotation and nonradial oscillations on the rapidly rotating Oe star ζ Ophiuchi (Walker et al. 2005).

Following the null detection on Procyon (Matthews et al. 2004), η Boo represents a critical target. For MOST, it is the first Sun-like star for which a positive detection of *p*-modes has been made.

In this paper we present precise frequencies and identifications of radial (*l* = 0) *p*-modes of low overtone *n* (where *n* represents the number of nodes in the eigenfunction between the center of the star and its surface) for η Boo based on MOST data, unhampered by aliasing. We describe our data reductions in the next section. A detailed mathematical description of the SigSpec procedure, which we use to obtain a statistical measure of the significance of the spectrum peaks, can be found in Reegen et al. (2005). In the third section of the paper we describe our efforts to model η Boo's *p*-mode frequencies. We show that the mode frequencies observed by MOST strongly constrain the mass, age, and composition of η Boo. We also show using the higher-overtone modes identified by K2003 combined with the MOST lower-overtone mode identifications that our stellar models, constrained solely by the oscillation data, fail to predict the structure of the outer envelope of η Boo correctly. In our conclusions, we motivate future observations of η Boo by the possibility of seeing *g*-modes in this star.

## 2. PHOTOMETRY

MOST observed η Boo for a total of 27 days (648 hours) from 2004 April 14 to May 11, inclusive. The integration time was 6 seconds, and the sampling rate was 3 times per minute. The readings were taken through a custom broadband optical filter (350 – 750 nm) which has about 3 times the throughput of a Johnson V band pass[2].

The overall duty cycle for this observing run was almost 96%; about 26 hours were lost in 3 gaps (lasting 10, 10 and 6 hours each). However, once one considers rejection of outliers due to severe radiation hits during passage through the South Atlantic Anomaly on some orbits, and intervals of high stray Earthshine, the net duty cycle is about 78%. These gaps typically last a few minutes each and are not spaced regularly in time. Hence, they do not degrade the spectral window of the time series, and the corresponding amplitude spectrum is virtually free of any one day alias sidelobes. The total number of data points in the time series analyzed in this paper is about 91,000.

---

[1] MOST is a Canadian Space Agency mission operated jointly by Dynacon, Inc., the University of Toronto Institute of Aerospace Studies, and the University of British Columbia, with assistance from the University of Vienna.

[2] For this analysis only MOST's Science Data Stream 2 (SDS2) was used; see Walker et al. 2003. SDS2 data consist of resolved images of the MOST telescope pupil illuminated by starlight, whereas SDS1 data are raw integrated photometric values preprocessed on board the satellite. Although including the SDS1 data increases the sampling rate and the total number of points in the photometry, they were rejected in this case because they reduce our ability to correct for stray light contamination and hence degrade the overall noise level.

Each SDS2 pupil image covers about 1500 pixels, and the image is almost fixed on the CCD detector even as the telescope pointing wanders slightly. These factors make the data relatively insensitive to guiding errors, but corrections are still applied for this effect and cosmic ray hits. A detailed description of the MOST processing procedures for the raw pupil image photometry are described in Reegen et al. (2005). Here we provide a summary of the important features of the reductions.

The main purpose of the data reduction is to reduce the effects of stray light from Earthshine by correlating pixels containing starlight and stray light with those containing only stray light. The technique has proven to be very successful, reducing the amplitudes in the Fourier domain of the orbit-modulated stray light contamination signal by at least two orders and in the best case by four orders of magnitude.

Each correlation step necessarily also reduces signal in the target pixels. This effect is due to the fact that – in case of pure noise – the slope of a linear regression in the intensity-intensity diagram will scatter about zero, and that a trend correction will always work in the sense of reducing both positive and negative deviations. Fortunately, this loss of signal can be determined from the amplitude decrease after each correlation step (which turned out to be constant and frequency independent), but can be also estimated from the decrease of the orbit harmonics amplitude. However, any tiny residual stray light contribution in the data is easily handled, since the orbit frequency ($\nu_{orbit}$ = 165 $\mu$Hz), its harmonics, and its daily side lobes due to a variable terrestrial albedo are well known. Such tiny residual orbit modulations do not affect, via the spectral window, the significance analysis.

The Fourier amplitude spectrum of the time series is shown in Figure 1 (a) and (b). The data have been stray light corrected and prewhitened with orbital (164.34 $\mu$Hz) and orbit-harmonic frequencies. But no trends have been removed nor offsets for subsets of data (corresponding to night-means in ground based photometry). This policy conserves the very low frequency effects seen in Figure 1 (a), which may be intrinsic, artifacts or of instrumental origin. We can live with this hitherto not investigated low frequency amplitudes because the spectral window of the MOST data is clean and close to a delta-function, as shown in the insert in Fig. 1 (a). Hence, they do not influence the frequency range of interest of this paper. Very low frequency peaks (below about 150 $\mu$Hz) are essentially found in every MOST dataset. Some components are associated with the CCD controller board temperature. But even when that is corrected for residual long-term variations, which can range in amplitude from a few 0.001 mag to 0.0001 mag, remain and do not repeat for other target stars. We found similar effects also in WIRE data and speculate that they are indeed intrinsic.

The amplitudes quickly fall off above 2000 $\mu$Hz to a noise level of approximately 10 micromagnitudes. Fig. 1 (b) shows the same spectrum zoomed in to a frequency range appropriate for p-modes on η Boo. Vertical lines show the location of the peaks we claim are radial p-modes (modes 3–10 in Table 1).

Compared to top quality ground based photometry the noise level is extremely small, at the level of about 10 parts per million. Unfortunately, the photometric amplitudes of p-modes are not expected to be significantly higher than this.

To identify significant peaks in the spectrum, we have applied a false-alarm probability approach. The routine SigSpec was developed by Reegen (2005) to compute the statistical probability that an amplitude peak at a given frequency *and phase* angle is not produced at random. The underlying probability density function of the amplitude spectrum generated by pure noise may analytically be deduced, only if phase-dependency is considered as well. We note that these dependencies are not considered, for example, by the popular Scargle-Lomb (1982) criterion. An analytically correct treatment of the systematic distortion of Fourier amplitudes due to the restriction to a finite time interval and the individual characteristics of the spectral window can only be obtained by the inclusion of frequency- *and phase-* dependencies.

SigSpec introduces the significance, $\sigma(A) = -\log \Phi_{FA}(A)$, of amplitude levels $A$ in the frequency domain with a false-alarm probability $\Phi_{FA}(A)$. Significance is the (logarithmic) number of noise datasets to be analysed on average to obtain an amplitude at least as high as A at a given frequency and phase. A significance level of 6, for example, means that the peak amplitude $A$ would arise at the frequency and phase of the peak by chance in one out of $10^6$ cases.

SigSpec does not take into account coloured noise. Since there are both instrumental effects (e.g., CCD readout, stability of spacecraft position) and stellar variations (e.g., low-amplitude modes, granulation noise) to be considered and neither of these two sources may be determined unambiguously, it presently is impossible to deduce a reliable noise amplitude spectrum for MOST. The heuristic approach to generate such a noise spectrum by means of (weighted) moving averages suffers from the presence of unresolved peaks which increases the noise level and the risk of missing intrinsic signal.

The significance peak spectrum generated by SigSpec from the amplitude spectrum in Fig. 1 (a) is shown in Fig. 1 (c), in which only peaks with significance ≥ 4.0 are plotted. At this threshold (4), only one out of ten thousand randomized data sets will accidentally yield the same Fourier amplitude and phase at a given frequency. Throughout this research we only consider modes with sig ≥ 6.9. We chose this threshold for several reasons. All of the radial order modes have sig ≥ 6.9. At higher thresholds, for example at sig ≥ 8.0, modes 8 and 9 from the radial sequence would have to be eliminated. This would not have impacted our model analysis since it is the modes with lower frequencies that set the most stringent constraints on the models. When we decrease the threshold below sig = 6.9 the echelle field becomes increasingly confused with peaks. Furthermore, as we show in section 3.8, the number of modes that appear in the echelle diagram with sig ≥ 6.9 is consistent with the total number of $l = 0$, 1, and 2 $p$-modes we expect to see in the frequency range 200 $\mu$Hz to 600 $\mu$Hz. A complete set of Fourier and SigSpec spectra are provided in the MOST Public Data Archive at www.astro.ubc.ca/MOST.

## 3. ANALYSIS

### 3.1 Mode Identifications

To help identify the oscillation peaks we plot the frequencies of our peaks with significance greater than or equal to 6.9 in an echelle diagram with folding frequency equal to 40 $\mu$Hz (see Fig. 2). The value for the folding frequency was chosen because it closely corresponds to the model predicted average large spacing between adjacent $p$-modes. The echelle diagram reveals a clear $l = 0$ sequence of modes from 200 $\mu$Hz to 500 $\mu$Hz, along with a scatter of other peaks. We have labeled (and encircled) the peaks that we believe could be $l = 0$ $p$-modes. Of the labeled modes, the sequence of modes labeled from 3 to 10 is the most certain. The five peaks that could, in some combination, represent the continuation of the $l = 0$ sequence to lower frequencies, labeled "1a," "1b,"  "1c," "2a," and "2b" in Fig. 2, are less certain. The two peaks, labeled "11" and "12" in Fig. 2 line up with the 3–10 sequence but for our following modeling analysis we consider only the 3–10 sequence of modes. The problem with modeling peaks 11 and 12 will be discussed later in section 3.5 in connection with K2003. We also look at reasonable extensions of the 3–10 sequence to lower frequencies. The frequencies of the labeled peaks are listed in Table 1. We do not see any evidence for other equi-spaced sequences that could be associated with $l = 1$ or $l = 2$ $p$-modes. The other points in Fig. 2 are more difficult to identify because, according to our models, the $l = 1$ and 2 $p$-modes are not expected to fall along equi-spaced sequences in this frequency range. These peaks are discussed later.

We show in Figure 3 the frequency peaks identified in K2003 and C2005 in an identically scaled echelle diagram to Fig. 2. The radial mode sequence becomes visible only after some of the peaks are corrected for 1/1d aliasing. Of the eight $l = 0$ modes in K2003 six were obtained by shifting the uncorrected peaks by 1/1d and of the eight $l = 0$ modes in C2005 four were obtained by shifting the uncorrected peaks by 1/1d. Because one has to decide which peaks should be shifted, and in which direction, some ambiguity in the mode identifications exists. Alias corrections are more difficult to judge in the case of nonradial modes where, because of mode bumping, the modes may not lie on a well defined line in the echelle diagram. In the case of η Boo, there is the added complication that the large frequency spacing, ~40 $\mu$Hz, is almost a multiple of the 1/1d alias correction, 11.57 $\mu$Hz. The advantage of the uninterrupted viewing from space is clear, *alias corrections are not required to interpret the peaks in the oscillation spectrum obtained from MOST.*

In Figure 2 we also plot, along with the MOST peaks, the alias corrected $l = 0$ modes from K2003 and C2005. The K2003 and C2005 modes pick up where the MOST modes drop away around 600 $\mu$Hz. In the region of overlap MOST mode 11, at 611.5 $\mu$Hz, is seen in both K2003 (at 611.0 $\mu$Hz) and C2005 (at 610.6 $\mu$Hz, which is an alias corrected peak) and MOST mode 12 at 650.4 $\mu$Hz is seen in K2003 (at 651.2 $\mu$Hz, which is an alias corrected peak). Furthermore, we note that the MOST $l = 0$ sequence joins smoothly with the $l = 0$ sequence from K2003 and C2005. Although both $l = 1$ and $l = 2$ sequences are also visible in the alias corrected K2003 data (see Fig. 9 in K2003) and the alias corrected C2005 data (see Fig. 6 in C2005) above 600 $\mu$Hz, no continuation of either of these sequences below 600 $\mu$Hz is discernable in the MOST data. This we show from our models is precisely what is expected and illustrates the inherent problem of arbitrary alias corrections. We cannot explain the slight discrepancy between the K2003 and C2005 mode frequencies above 700 $\mu$Hz.

### 3.2 Introduction to Model Analysis

We choose to focus our model analysis at this time on the radial modes because their identification as noted in the previous section is less ambiguous. Unless forced by inconsistencies to assume otherwise, we presume the oscillation data are real and the models are correct. We attempt to fit model oscillation spectra to the labeled MOST peaks, to the $l = 0$ K2003 alias corrected peaks shown in Fig. 2 and 3, and to the combined MOST plus $l = 0$ K2003 peaks. We are using the K2003 peaks, instead of the C2005 peaks, only because we have already modeled the $l = 0$, 1 and 2 peaks from K2003 as reported in Guenther (2004). In the summary and conclusions we will compare our model results with those of C2005. The MOST frequencies are known with an uncertainty of approximately ±0.4 $\mu$Hz and the K2003 frequencies with an uncertainty of approximately ±1 $\mu$Hz.

Our model analysis is similar to that described in detail in Guenther (2004), which was originally motivated by the global optimization techniques used by Fontaine and his coworkers to analyze the nonradial oscillation spectrum of hot subdwarfs (Brassard, Fontaine, & Billères et al. 2001, Charpinet et al. 2005). We search a dense and extensive grid of stellar models for close matches between the observed and model oscillation spectra. The quality of the match is quantified by the simple chi-squared relation,

$$\chi^2 \equiv \frac{1}{N} \sum_{i=1}^{N} \frac{(\nu_{obs,i} - \nu_{mod,i})^2}{\sigma_{obs,i}^2 + \sigma_{mod,i}^2},$$

where $\nu_{obs,i}$ is the observed frequency for the i$^{th}$ mode, $\nu_{mod,i}$ is the corresponding model frequency for the i$^{th}$ mode, $\sigma_{obs,i}$ is the observational uncertainty for the i$^{th}$ mode, $\sigma_{mod,i}$ is the model uncertainty for the i$^{th}$ mode, and $N$ is the total number of matched modes. Details of the searching and mode matching procedure are described in Guenther and Brown (2004). The model uncertainty, estimated from fits of the solar $p$-mode frequencies to standard solar models, increases from a hundredth of a percent to half of a percent as the frequencies of the modes approach the acoustic cut-off frequency (see Fig. 1, Guenther & Brown 2004). We stress that our estimate of the uncertainties in the model is at best an educated guess. We do not know how the uncertainties scale with mass, age, and composition since we only have the Sun with which to compare. Since the model uncertainties in the frequency range observed by MOST are an order of magnitude smaller than the observed frequency uncertainties they do not have a significant impact on the evaluation of $\chi^2$.

For our analysis of η Boo we consider grids of models with mass fraction of hydrogen $X = 0.69$ and 0.71; mass fraction of metals $Z = 0.02$, 0.03, and 0.04; masses ranging from 1.400 $M_\odot$ to 1.900 $M_\odot$ in steps of 0.005 $M_\odot$, and evolutionary age ranging from the zero-age main-sequence to the base of the giant branch (resolved by approximately 1000 models). For the actual search, the oscillation spectra of the models in the grid are interpolated by a factor of 10 in age and mass. In total approximately 300,000 models were used and 100 times that number of interpolated oscillation spectra were compared to the observed spectrum of η Boo.

The models were constructed using the Yale Stellar Evolution Code (YREC, Guenther et al. 1992). The physics of the models, described in Guenther and Brown (2004), are current and include: OPAL98 (Iglesias and Rogers 1996) and Alexander and Ferguson (1994) opacity tables, Lawrence Livermore equation of state tables (Rogers 1986; Rogers, Swenson and Iglesias 1996), and nuclear reaction cross sections from Bahcall et al. (2001). The mixing length parameter, an adjustable parameter which sets the temperature gradient in convective regions according to the Böhm-Vitense (1958) mixing length theory, was set from calibrated solar models constructed using the same input physics. The model pulsation spectra were computed using Guenther's nonradial nonadiabatic stellar pulsation program (Guenther 1994). The code uses the Henyey relaxation method to solve the linearized nonradial nonadiabatic pulsation equations. The nonadiabatic component includes radiative energy gains and losses as formulated in the Eddington approximation but does not include coupling of convection to the oscillations (see Balmforth 1992 and Houdek et al. 1999 for discussion of the effects of convection on oscillations).

To visualize how the oscillation modes constrain the models consider the example in Fig. 4 where we show $\chi^2$ versus mass and age for models with $X = 0.71$, $Z = 0.04$ and for the MOST $l = 0$ p-modes between 200 μHz and 500 μHz (i.e., the 3-10 sequence of frequencies in Table 1). Only the lowest values of $\chi^2$ are shown. There is a clearly defined minimum around 2.3 Gyr and 1.7 M$_\odot$. This, of course, corresponds to the models whose oscillation spectra most closely match the observed spectrum according to our definition of $\chi^2$. Since no other constraints are applied to the model, the model is constrained exclusively by the observed oscillation spectrum. To facilitate comparisons with models with other compositions we select the bottom edge of $\chi^2$ values, i.e., we select the minimum $\chi^2$ value for each mass, and plot this resultant curve in projection on 2D plots of $\chi^2$ versus mass. Even though we are showing in projection the minimization of $\chi^2$ as a function of mass, the age of the model is also constrained since the constrained models lie on a nearly flat two dimensional curve in the $\chi^2$ versus mass and age plot.

### 3.3 Model Analysis of MOST peaks

In Figure 5 we show $\chi^2$ versus mass for the six combinations of $X$ and $Z$. We also plot, in Fig. 5, $\chi^2$ for fits to the adiabatic frequencies for the models with $X = 0.71$. Note that the $\chi^2$ values running along the upper right-hand side of the plot correspond to model fits in which the spacing between adjacent p-modes in the model is approximately 20 μHz, one-half that expected for η Boo. We ignore these model fits since they are positioned far from η Boo's location in the HR-diagram. The $\chi^2$ values determined from the nonadiabatic frequencies are slightly lower than for the adiabatic frequencies suggesting that the nonadiabatic calculation is indeed an improvement over the adiabatic calculation. This has previously been demonstrated in Guenther (2004) for η Boo using K2003 observations and in Guenther and Brown (2004) for the Sun and α Cen A.

To place the $\chi^2$ values in perspective, consider that they are calculated assuming that the MOST frequencies are accurate to ±0.4 μHz. Note that the intrinsic model uncertainties $\sigma_{mod}$ in this n-value range are estimated from models of the Sun to be an order of magnitude smaller, hence, do not affect significantly the $\chi^2$. Therefore, a $\chi^2 = \sim 1$ implies that the model mode frequencies are within one sigma or ~0.4 μHz of the observed frequencies, a $\chi^2 = \sim 4$ implies the model mode frequencies are within two sigma of the observed frequencies, etc. For η Boo, $\chi^2 = \sim 1$ means all of the model frequencies match the observed frequencies to within ~0.1%.

The n-values (the radial order of the mode) of the best fitting model to the 3-10 MOST modes are 3 to 10 inclusive, the same value as the ID (a coincidence only). We stress that the n-value of the mode is not assumed when matching observed modes to model mode frequencies. The MOST observed p-modes are of much lower radial order than the p-modes identified by K2003 and C2005. As we discuss later, we speculate that MOST might be able to see even lower order p-modes, and possibly g-modes.

The mass of the model with minimum $\chi^2$ depends sensitively on composition. Table 2 lists the mass, age, log $T_{eff}$, and log L/L$_\odot$ for the models at the $\chi^2$ minima, in Fig. 5, labeled "3-10." The mass of the best fitting model (i.e., the model whose oscillation spectrum most closely matches the observed spectrum) increases as the assumed $Z$ of the model is increased. This was previously noted in Guenther (2004) and in C2005. We also note that decreasing $X$ decreases the mass of the best fitting model.

We plot, in Fig. 6, the HR-diagram positions of the models with minimum $\chi^2$ for all the compositions considered. Following Di Mauro et al. (2003) we take the luminosity of η Boo to be L/L$_\odot$ = 9.02 ± 0.22 (log L/L$_\odot$ = 0.955±0.01) and its effective temperature to be $T_{eff}$ = 6028 ± 45 K (log $T_{eff}$ = 3.780±0.003). Di Mauro's luminosity is based on the Hipparcos parallax π = 88.17±0.75 mas and his effective temperature is obtained from the average of several published determinations. Two evolutionary tracks are plotted to help visualize the evolutionary phase of the models and η Boo. The model data points are labeled 'zxx' for $Z = z/100$ and $X = x/100$. The $X = 0.71$ and $Z = 0.04$ model (labeled '471') lies closest to η Boo, almost within one sigma of the uncertainties in log $T_{eff}$ and log L/L$_\odot$. We discuss the other unlabeled data points in the plot later in section 3.5.

The $\chi^2$ plots and the HR-diagram plot both suggest that $Z = 0.04$ is optimum (of the 3 metallicities considered). The $\chi^2$ plots favor $X = 0.71$ over $X = 0.69$ models but the distinction is not great. Although the best fitting models to the oscillation data lie close to the observed position of η Boo in the HR-diagram, they are not coincident. The uncertainties in η Boo's HR-diagram position could be underestimated, especially log $T_{eff}$ (see C2005) or, the models could have too small radii (which would shift the models slightly to cooler temperatures). The latter possibility would be the result of inaccurate modeling of the outer layers, specifically the convective envelope.

We return to this subject when we analyze the MOST and K2003 modes combined.

In Fig. 7 we plot in an echelle diagram all the MOST frequency peaks with significance ≥ 6.9 (similar to Fig. 2) along with the frequencies of the modes of the best fitting model (i.e., the $Z = 0.04$, $X = 0.71$ model). The adiabatic and nonadiabatic radial $p$-mode frequencies, the nonadiabatic $l = 1$ and $l = 2$ $p$-mode frequencies, and the $l = 1$ and $l = 2$ $g$-mode frequencies for the model are all shown. Data which line up along a vertical path in this diagram correspond to modes whose frequencies are spaced regularly by ~40 $\mu$Hz.

Following along the solid line, which represents the $l = 0$ (radial) model $p$-modes, we see that the line passes near or through 8 of the filled circles between 200 $\mu$Hz and 500 $\mu$Hz. These 8 filled circles are the MOST radial modes that were used to constrain the model fit. The model and the 8 MOST modes are consecutive radial modes. The model mode frequencies match the MOST frequencies to better than 1 $\mu$Hz, typically 0.5 $\mu$Hz.

The solid line resolves into two distinct lines above 800 $\mu$Hz with the left line corresponding to nonadiabatic model frequencies and the right line corresponding to adiabatic frequencies. The 8 MOST peak frequencies are fit almost equally well by the adiabatic and nonadiabatic model frequencies, consistent with the fact that the adiabatic and nonadiabatic $\chi^2$ curves (Fig. 5) are not very different.

The $l = 1$ model modes (upward pointing triangles) loosely fall along a vertical path near folded frequency =35 $\mu$Hz but this alignment becomes increasingly disorganized at lower frequencies. The $l = 2$ model modes (downward pointing triangles) follow a path nearly parallel to the $l = 0$ modes but this path also becomes increasingly disorganized at lower frequencies. The model frequencies of the $l = 1$ $p$-modes below 600 $\mu$Hz and the $l = 2$ $p$-modes below 400 $\mu$Hz are heavily bumped and, as a consequence, do not lie along well defined paths.

We also note the existence of $g$-modes in the frequency range 150 $\mu$Hz to ~370 $\mu$Hz in the model. These modes are mixed modes with $n_G$, the number of nodes with $g$-mode–like character (Scuflaire 1974), ranging from 1 to 80. In the same region of the echelle diagram that the models predict $g$-modes we see a large number of peaks in the MOST spectrum.

We discuss the nonradial modes at greater length in section 3. 7.

### 3.4 The lowest frequency l = 0 MOST Peaks

We are motivated to try to discern which, if any, of the peaks below 200 $\mu$Hz (peaks 1a, 1b, 1c, 2a, and 2b in Table 1) are also $l = 0$ $p$-modes because the lowest $n$-valued $p$-modes provide the strongest constraints on the interior structure of our stellar models. We cannot tell directly if they are $p$-modes because they lie within a frequency range for which the MOST instrument was not optimized during this particular observing run. In this region, many real modes, including possibly $g$-modes are likely crowded together with peaks of instrumental origin. We compare the $\chi^2$ curves of several possible combinations of the 1 and 2 peaks combined with the above 200 $\mu$Hz peaks (3–10). In Fig. 8 we plot $\chi^2$ for: the 3–10 peaks; the 2a and 3–10 peaks; the 2b and 3–10 peaks; the 1a, 2b, and 3–10 peaks; the 1b, 2b, and 3–10 peaks; and the 1c, 2b, and 3–10 peaks. The $\chi^2$ curves, in Fig. 8, are all computed for $X$ =0.71 and $Z$ =0.04 models. Other compositions have been computed but are not shown here since they duplicate the relative behavior shown in Fig. 8. We see that the $\chi^2$ curves become narrower as more modes are included. Of the two sequences extended by either the 2a or 2b peak, the $\chi^2$ curves dip to lower values for the sequence extended by the 2b peak. Based on the observed oscillation spectrum alone and the assumed validity of the stellar modeling we conclude that the 2b peak is more probably an $l = 0$ $p$-mode than the 2a peak. Similarly we conclude that either the 1a or 1b peak is more likely to be an $l = 0$ $p$-mode than the 1c peak, with the difference between the 1a and 1b peaks being minimal. We emphasize that the $\chi^2$ method cannot be used to prove that any of these peaks are radial modes. We are only showing there exist standard stellar models with oscillation spectra that fit the 1b and 2b peaks combined with the 3-10 peaks better than other combinations of peaks.

The properties of the 1b, 2b, 3-10 best fitting models are listed in Table 2. The surface temperature and luminosity of the $Z$ =0.04 models lie within two sigma of η Boo. We note that the surface temperature and luminosity of the best fitting models to the other mode combinations lie many sigma farther away from η Boo.

These tests show that the dense grid and $\chi^2$ method is indeed very sensitive to low-$n$ $p$-mode frequencies and if, in the future, we can ascertain the validity of these modes, they will provide strong constraints on the structure of η Boo's interior.

*3.5 Testing the Model Fits at High n-Values.*

In this section we examine and compare, using echelle and HR diagrams, models that fit: 1. the MOST 3-10 modes, 2. the K2003 $l = 0$ modes, and 3. the combined MOST 3-10 modes and K2003 $l = 0$ modes. As one goes to higher frequencies (equivalently, higher $n$) the frequencies of the modes become increasingly more sensitive to the outer layers. In the case of the Sun, there is a well known discrepancy between the model frequencies and the observed frequencies which is most often recognized as a failure of the mixing length theory to accurately describe the temperature structure of the superadiabatic layer. To see if a similar discrepancy exists for η Boo we add to the lower frequency MOST modes the higher frequency radial modes of K2003. It has to be stressed that the first two modes of K2003 coincide well within the measuring errors with MOST modes "11" and "12", hence supporting with independent data and analyses the K2003 identification of $l = 0$ p-modes. These two modes are also seen in C2005. For consistency reasons we are using only the radial frequencies beyond 600 $\mu$Hz in K2003.

In Fig. 9, we plot an echelle diagram for the MOST modes, the radial K2003 modes, and the modes from several different model fits. The model fits were obtained in the same way as described in section 3.3 for the 3-10 MOST modes. Here we only show models with $Z = 0.04$, $X = 0.71$. Other compositions do not fit as well (see Table 2).

We first look at the model fits labeled "K" in Table 2, which correspond to the best fitting models to the $l = 0$ K2003 p-modes only. Note that the best fit model at $X = 0.69$, $Z = 0.02$ was below the 1.4 M$_\odot$ lower edge of the grid used in this study, hence, is not included in the table. Compared to the 3-10 model fits, the K model fits are skewed to lower masses. In the echelle diagram, Fig. 9, we see the nonadiabatic K model frequencies fit the radial K2003 modes well but the fit is outside the observational uncertainties ($\pm 0.4$ $\mu$Hz) for MOST modes 3, 4, and 7.

In Fig. 6, we plot the HR-diagram positions of the best fitting models along with the position of η Boo. The large plot symbols correspond to the best-fit models, i.e., the models with minimum $\chi^2$. The two diagonal lines of small plot symbols extending from the MOST 3-10 best-fit models to the K2003 $l=0$ best fit models correspond to all the models that fit the K2003 $l=0$ with $\chi^2 \leq 1$. First consider the top right line of small plot symbols. Each point corresponds to a model fit, with $Z = 0.04$ and $X = 0.71$, to the K2003 $l = 0$ mode frequencies that has $\chi^2 \leq 1$. The large plot symbol locates within this sequence the model with the minimum $\chi^2$. The second line of small plot symbols corresponds to a model fit, this time with $Z = 0.02$ and $X = 0.71$, but again to the K2003 $l = 0$ mode frequencies that has $\chi^2 \leq 1$. Although one might have expected an ellipse-like distribution of models, the oscillation frequencies actually constrain the models to an extremely narrow ellipse, too narrow to be resolvable on the scale of the plot shown. Essentially, the oscillation frequencies constrain the models to a unique linear relationship between log $T_{\rm eff}$ and log L/L$_\odot$ (see Guenther 2004 for a general discussion of this behavior). What these sequence of points show is that a large range of models can fit the higher frequency modes identified by K2003 within one sigma of the observed frequency uncertainties. This point should be kept in mind when comparing model fits to higher frequency radial modes of K2003 and C2005.

It is important to note that the line of models constrained by the K2003 $l = 0$ with $\chi^2 \leq 1$ passes through the MOST 3-10 best fit models, and also the MOST 3-10 plus K2003 $l = 0$ best model fits. This demonstrates that the K2003 and MOST models are consistent with each other within the uncertainties of the observations. That is, with in the uncertainties of the observations, model fits to MOST's modes lie within one-sigma of the model fits to the K2003 $l = 0$ modes. Because the MOST frequencies are lower, they provide a much stronger constraint on the interior of the model, and, indeed, within the uncertainties uniquely constrain the model in the HR-diagram. This also shows that the commonly adopted approach of "finding" a spectrum that fits models already constrained by the star's observed position in the HR-diagram can be misleading. As Fig. 6 shows, it is possible to find one-sigma fits to the K2003 observations and the observed position in the HR-diagram, but this fit ignores the fact that many other models, including those that fall far from the observed position in the HR-diagram, also fit the observations within one-sigma. By quantifying the quality of the fit to the oscillation data we avoid over interpreting the significance of a model fit.

We next consider the combined MOST and K2003 modes. For the MOST 3-10 model fit, discussed in section 3.3, the match between the observed radial modes and the model radial modes systematically worsens with increasing frequency. To see this consider the model fits to the MOST plus K2003 modes. For the MOST 3-10 plus K2003 modes, labeled "3-10 + K" in Table 2, the fits lie between the K and the 3-10 model fits as seen in both the HR-diagram plot (Fig. 6) and the echelle diagram (Fig. 9). Although the echelle diagram fit appears to be OK, the HR diagram position of the model is far from η Boo's observed position. There is an inconsistency caused by either problems in the modes, the models, or the HR-diagram position of η Boo. We suggest that the MOST frequencies are constraining the interior of the model correctly and that the discrepancy at higher frequencies is due to the failure of the model to accurately predict the structure of the surface layers. This speculation is consistent with similar conclusions drawn from fits of solar models to the Sun's observed $p$-mode frequencies. The possibility that the HR-diagram position of η Boo is wrong and should be near where the combined MOST plus K2003 model fits indicate is less likely since it would require a four sigma correction in the corresponding observations. The consistency between the K2003 and C2005 modes themselves, the consistency of model fits to the K2003 and MOST modes, and the coincidence of two modes in common between MOST, K2003 and C2005, all suggest that the mode identifications are correct. Needless to say, confirming $p$-mode observations would significantly strengthen the case.

In summary we conclude that the discrepancy in frequencies between the model and the observations at higher $n$ is probably real. The low-$n$ $p$-modes provide the strongest constraints on the interior structure of the model and effectively anchor the model. The poorer fit at higher frequencies suggests the model for the outer layers is inaccurate.

### 3.6 Evolving p-modes

To demonstrate how well the model frequencies match the MOST observed frequencies we show in Fig. 10 a plot of evolving $p$-modes and the 1b, 2b, 3-10 MOST peaks. The $l = 0$ (points and lines) and $l = 1$ (points only) $p$-mode frequencies for models with $M = 1.70$ $M_\odot$, $Z = 0.04$, and $X = 0.71$ are plotted as a function of the age of the model. The 1b, 2b, 3-10 MOST peaks are plotted at age = 2.40259 Gyr (see Table 2), the age where the observed frequencies (3-10) best match the model.

We note that the $l = 1$ $p$-modes of the models do not fall half-way between the $l = 0$ $p$-modes. As the star evolves the frequencies of the lowest order nonradial $p$-modes are perturbed by the occurrence of $g$-modes with similar frequencies (i.e., mode bumping or avoided crossings). At the age of our model fit to η Boo, the lowest order $l = 1$ $p$-mode frequencies lie above 350 $\mu$Hz. The broken $l = 1$ segments below 350 $\mu$Hz correspond to mixed $g$-modes where the number of $g$-mode nodes is greater than the number of $p$-mode nodes (Scuflaire 1974; Unno et al. 1989). This suggests that the peak frequencies identified by MOST below 350 $\mu$Hz (see Fig. 7), which are not radial modes and not of instrumental origin, could well be nonradial, mixed-mode $g$-modes. Above 350 $\mu$Hz we do not expect the $l = 1$ $p$-modes, in an echelle plot, to lie along a well defined sequence of equally spaced modes, similar to the $l = 0$ modes.

### 3.7 Sensitivity of Nonradial p-modes to Mass

We have stated that the low frequency nonradial modes will be difficult to identify because their frequencies are strongly affected by mode mixing (avoided crossings). To convey the difficulty we show in Fig. 11 an echelle diagram of the $l = 0$, 1 and 2 $p$-mode frequencies for two models for η Boo that are nearly identical. In the background we also show the frequency peaks from MOST with sig ≥ 6.9. The model frequencies range from $n = 1$ upward. The $M = 1.710$ $M_\odot$ model has log $T_{eff} = 3.7838$, log $L/L_\odot = 0.946$ and the $M = 1.715$ $M_\odot$ model has log $T_{eff} = 3.7849$ and log $L/L_\odot = 0.952$. The radial modes ($l = 0$ panel) show only a small perturbation in frequency between models. The nonradial modes, on the other hand, show a relatively large perturbation in frequency between models for some of the modes, especially the $l = 1$ $p$-modes.

Even though we have already constrained our model of η Boo reasonably well using radial modes only, the fit is not precise enough to enable us to easily fit the nonradial modes by simply perturbing this best fit model. Small perturbations in mass lead to, in some cases, large perturbations in mode frequencies.

Our dilemma is compounded for MOST data since only the lower frequency modes in η Boo are seen. According to models at higher frequencies the mode bumping effects diminish enough to enable an $l = 1$ and $l = 2$ sequence to be seen. But consequently, this higher frequency domain (accessible from ground by spectroscopy) does not have the same diagnostic power as the lower frequency domain (accessible by photometry). It is the same sensitivity which does not yet allow us to unambiguously claim which of the frequencies detected in the 2004 MOST photometry are intrinsic or perhaps of instrumental origin. For this purpose we need further observations. We cannot just look at our power spectrum and immediately identify the characteristic picket fence pattern associated with regularly spaced *p*-modes. We have had to use the echelle diagram to pick out the radial modes from the apparent scatter of nonradial modes.

### 3.8 Distribution of oscillation peaks

Although we cannot prove whether or not the peaks seen by MOST are nonradial modes, we can offer a statistical assessment of the number of peaks seen by MOST. In Fig. 12 we show the number of peaks seen by MOST, with sig ≥ 6.9, and the number of $l = 0, 1,$ and 2 *p*-modes and *g*-modes predicted by the $M = 1.710$ M$_\odot$ model. The bin size of the histogram is 80 $\mu$Hz. The histograms match from 200 $\mu$Hz to 900 $\mu$Hz. The number of modes below 400 $\mu$Hz in the model spectrum increases because of the inclusion of *g*-modes, which overlap the low frequency *p*-mode spectrum. That the number of MOST peaks also increases does suggest that MOST may be seeing *g*-modes. The MOST histogram drops abruptly at ~900 $\mu$Hz while the model histogram falls off at ~1100 $\mu$Hz, which is near the acoustic cutoff frequency for the model. We cannot explain the drop-off in MOST sensitivity above 900 $\mu$Hz. We do know it is not an artifact of SigSpec.

The MOST histogram above 1100 $\mu$Hz contains ~3 peaks every 160 $\mu$Hz. Since we do not at this time expect to see any modes above the acoustic cut-off frequency, this gives us some idea of how my false peaks, i.e., of non-stellar origin, are expected in the MOST spectrum for η Boo. If we assume in the worst case that the distribution of false peaks is flat then of the 56 peaks detected by MOST between 240 $\mu$Hz and 700 $\mu$Hz, with sig ≥ 6.9, only 9, i.e., 3 peaks per 160 $\mu$Hz, are likely to be false. In other words, more than 80% of the peaks (with sig ≥ 6.9) in the range where we identify *p*-modes are likely to be of stellar origin.

Until we can identify the nature of all of the peaks in the frequency range above 200 $\mu$Hz for η Boo we will always be left with the possibility that some of our radial mode identifications are incorrect. We hope that by re-observing η Boo, we will be able to confirm the stellar origin of all of the peaks (radial and nonradial) in the amplitude spectrum.

## 4. SUMMARY, DISCUSSION, AND CONCLUSIONS

We have analyzed the asteroseismic observations of η Boo obtained from MOST by comparing the oscillation spectra of Guenther's grid of stellar models to the observed frequency spectrum of peaks. We are able to identify eight consecutive radial *p*-modes of low *n*, with frequencies ranging from 210 $\mu$Hz to 500 $\mu$Hz and spacings of ~40 $\mu$Hz. The spectrum peaks do not require any 1/1d alias corrections. The MOST *p*-modes join up smoothly with the higher *n*-valued radial *p*-modes of K2003 (and C2005) that have been corrected for 1/1d aliasing.

Within Guenther's grid of stellar models we examined approximately 300,000 models with masses ranging from 1.4 M$_\odot$ to 1.9 M$_\odot$, ages ranging from zero-age main-sequence to the base of the giant branch, $X = 0.69, 0.71,$ and $Z = 0.02, 0.03,$ and 0.04. The model whose oscillation spectrum best fits the observed modes (3–10), with $\chi^2 = 1.4$ (which corresponds to the frequencies of the modes agreeing to within ~0.1%) , has a mass $M = 1.71 \pm 0.05$ M$_\odot$, age = 2.40±0.03 Gyr, $X = 0.71$, and $Z = 0.04$. The best-fit model's surface temperature, log $T_{eff} = 3.784$ and the luminosity, log L/L$_\odot$ = 0.9463 are close to those proposed by Di Mauro et al. (2003) who quote log $T_{eff} = 3.780 \pm 0.003$ and log L/L$_\odot$ = 0.955±0.01 based on weighted values taken from the literature. The estimate of uncertainty in our mass and age is based on the variation in mass and age of the best fitting models with respect to the hydrogen abundance variation ($X = 0.71 \pm 0.02$). The frequencies of the model fit the MOST 3-10 modes within observational uncertainty (±0.4 $\mu$Hz). Extending the observed frequency list to include the radial modes of K2003 shows that the difference, model frequency minus observed frequency, increases from less than ~0.5 $\mu$Hz at 250 $\mu$Hz to ~5 $\mu$Hz at 1000 $\mu$Hz. The difference is within observational uncertainty up to 700 $\mu$Hz.

We compared model fits to the MOST radial *p*-modes, which cover the frequency range 210 $\mu$Hz to 500 $\mu$Hz, combined with the alias corrected $l = 0$ *p*-modes of K2003, which cover the frequency range 600 $\mu$Hz to 1000 $\mu$Hz. We note that two MOST modes (which were not included in any of our model analysis), modes 11 and 12, do coincide within the uncertainties of the two lowest frequency modes of K2003, hence lending strong support to the $l = 0$ *p*- mode identifications in K2003. Our best fit model for the combined set of modes, with $\chi^2 = 2.3$, has a mass $M = 1.66 \pm 0.06$ M$_\odot$, age = $2.71 \pm 0.04$ Gyr, $X = 0.71$, and $Z = 0.04$. Significantly, though, the HR-diagram position of this model fit is over four sigma away from η Boo's position. Either the model is inaccurate or the η Boo HR-diagram position is wrong.

If we assume that η Boo's HR-diagram position is well determined and the MOST modes and K2003 modes are real, then the discrepancy at higher frequencies strongly suggests that the model inaccuracies are located in the outer layers. Although the temperature structure in the deepest layers of convective envelopes is well predicted by mixing length theory, the superadiabatic peak near the surface is not (i.e., its location, its height, and its width depend sensitively on the model of convection). The inaccurate temperature structure in the superadiabatic layer of our model affects the $p$-mode frequency predictions of all our modes, especially those modes that are sensitive to the outer layers, which for the low-$l$ modes observed on stars corresponds to the higher frequency modes. The discrepancy is similar to that seen in model fits to the Sun's oscillation spectrum. For the Sun, the discrepancy can be eliminated by incorporating 3D numerical convection results into the model (Rosenthal et al. 1999; Robinson et al. 2003). The Yale Convection Group and some members of the MOST science team are now working toward producing a 3D numerical simulation of the convective envelope of η Boo which we will incorporate in our best fitting model to see if the discrepancy can be eliminated or reduced. Preliminary results using convection simulations of the Sun scaled to η Boo look very promising (Straka et al. 2005).

Only by combining the low-$n$ modes obtained by MOST and the higher-$n$ modes of K2003 have we been able to study the deficiencies in our models. The MOST data set by itself does not extend to high enough frequencies where the discrepancy occurs. The K2003 data set by itself does not have any low frequency modes to anchor the model fits. By themselves, the $l = 0$ K2003 data can only weakly constrain the interior structure.

Although we know that our best fit model parameters do depend on input physics, for which we have intentionally kept as standard as possible, our model results compare favorably with the model results of C2005. C2005 use both their oscillation data AND the observed luminosity, effective temperature, and [Fe/H] to constrain their models, hence, their models by definition should fall within η Boo's observed position in the HR-diagram. C2005's best fit model has a mass of $1.57 \pm 0.07$ M$_\odot$ and our best fit model to K2003 data, with similar $X$ and $Z$ to that assumed in C2005, has a mass of $1.54 \pm 0.05$ M$_\odot$. Our best fit model to MOST modes has a slightly higher mass, primarily due to the higher $Z$ and $X$ of the model. The dependence of the model fits on mass and $Z$ has already been noted in Guenther (2004) and C2005. We emphasize that the lower frequencies of the MOST modes enables us to use the oscillation data by itself to constrain the star's HR-diagram position as well as its composition.

Below 210 $\mu$Hz we examined 5 peaks that could also be radial $p$-modes. We computed $\chi^2$ curves for several combinations of these modes and identified two of these lower frequency modes as possible radial $p$-modes, i.e., combinations that yielded the lowest values of $\chi^2$. This method was used to suggest lower frequency $p$-modes and to demonstrate the sensitivity of the models to low frequency modes. Additional observations are needed to confirm whether or not the low frequency modes are truly radial $p$-modes.

MOST sees many peaks in a region where $g$-modes are expected, between 200 $\mu$Hz and 350 $\mu$Hz. We are not at this time prepared to identify the peaks as $g$-modes because the region is also a region where with the current observational run we expect to have peaks of instrumental origin. Regardless, we speculate that because η Boo's convective envelope is thin, providing a shallow evanescent region for $g$-modes below to tunnel through, it could allow $g$-modes (and low-$n$ $p$-modes) to be seen at the star's surface. We are anxious to identify $g$-modes in the oscillation spectrum of η Boo because $g$-mode frequencies are sensitive to the helium core size and to the extent of convective core overshoot, the latter of which remains a critical and unknown parameter of great interest, especially in stellar isochrone research.

MOST also sees peaks scattered throughout the echelle diagram between 350 $\mu$Hz and 600 $\mu$Hz. Our models show that in this frequency range the nonradial $p$-modes are mixed (radial modes are not affected). Because the nonradial $p$-modes are mixed, their frequencies are heavily bumped from the more regularly spacing seen in $p$-modes at higher frequencies, and the modes do not fall along well defined vertical paths in the echelle diagram. The $l$-value of the modes are, as a consequence, difficult to determine. In addition, it is also difficult to fit stellar models to these low frequency nonradial modes since small adjustments to the model parameters lead to relatively large changes to the frequencies of the bumped modes. In order to establish, with reasonable certainty, that the peaks are of stellar origin MOST has re-observed η Boo in 2005. If we can identify recurring peaks as nonradial modes, we will be in a position to seriously test our models. Also, since we now know how to eliminate some of the sources of instrumental noise which contaminate below the 200 $\mu$Hz region of the spectrum, we hope to be able to see $g$-modes on η Boo.

We are left with many questions and much speculation. Why do η Boo's low $n$-valued $p$-modes have such (relatively) large amplitudes in luminosity? Why don't we see significant $p$-mode peaks above 650 $\mu$Hz in the MOST data for η Boo? Are some of the low frequency nonradial peaks seen by MOST $g$-modes? What is the best strategy to deal with mixed modes where both their identification and model fitting is difficult? Initial 3D numerical convection model results on η Boo show that its superadiabatic layer is located deeper into the star than in Procyon (Demarque 2005, private communication). Could it be that radiative damping, which is an issue in Procyon because its superadiabatic layer is very close to the surface, is less important in η Boo and that this is why MOST sees $p$-modes on η Boo and not on Procyon?

It is reassuring to our mission and those that follow, e.g., COROT (Baglin et al. 2002) that $p$-modes can be seen from space. We cannot help but be enthusiastic and encouraged by MOST's first asteroseismic results on η Boo. They have yielded results that confirm the stellar model, provide new constraints on models of stellar convection, demonstrate the complementarities of space photometry, and raise new questions for future missions to address.

The Natural Sciences and Engineering Research Council of Canada supports the research of D.B.G., J.M.M., A.F.J.M., S.M.R., G.A.H.W., A.F.J.M. is also supported by FQRNT (Québec), and R.K. is supported by the Canadian Space Agency. T.K, P.R. and W.W.W. are supported by the Austrian Research Promotion Agency (FFG), and the Austrian Science Fund (FWF P17580). We thank graduate students Chris Cameron and Jason Rowe for providing valuable input on the Fourier spectrum of η Boo.

FIGURE CAPTIONS

Fig. 1 (a) —The amplitude spectrum of η Boo from stray light corrected and with orbital (164.34 μHz) and orbit-harmonic frequencies prewhitened MOST data. The spectral window function is shown in the insert.

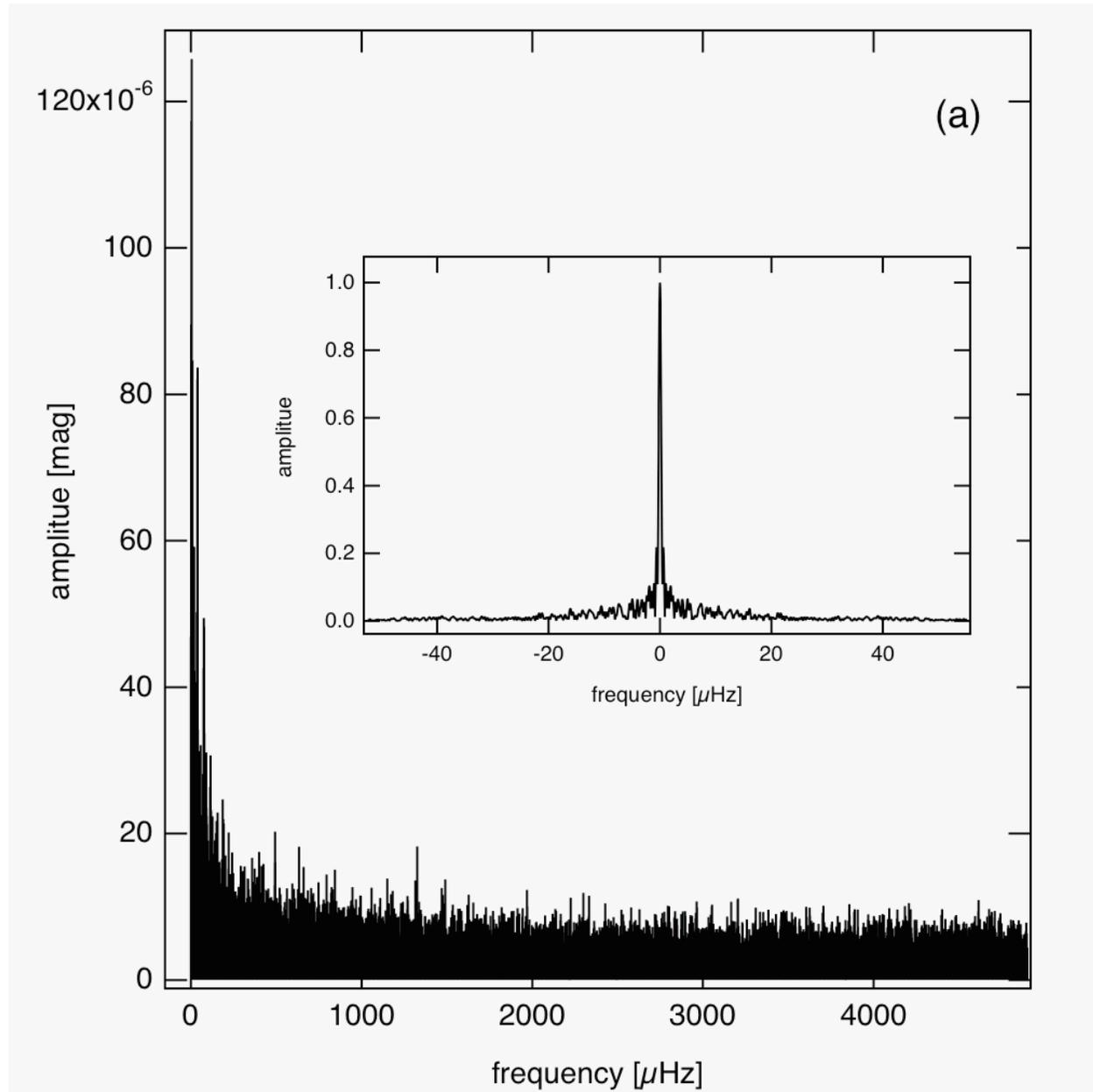

(b) Shows the amplitude spectrum within the range where *p*-modes are expected. Vertical lines indicate where the MOST 3–10 peaks are located.

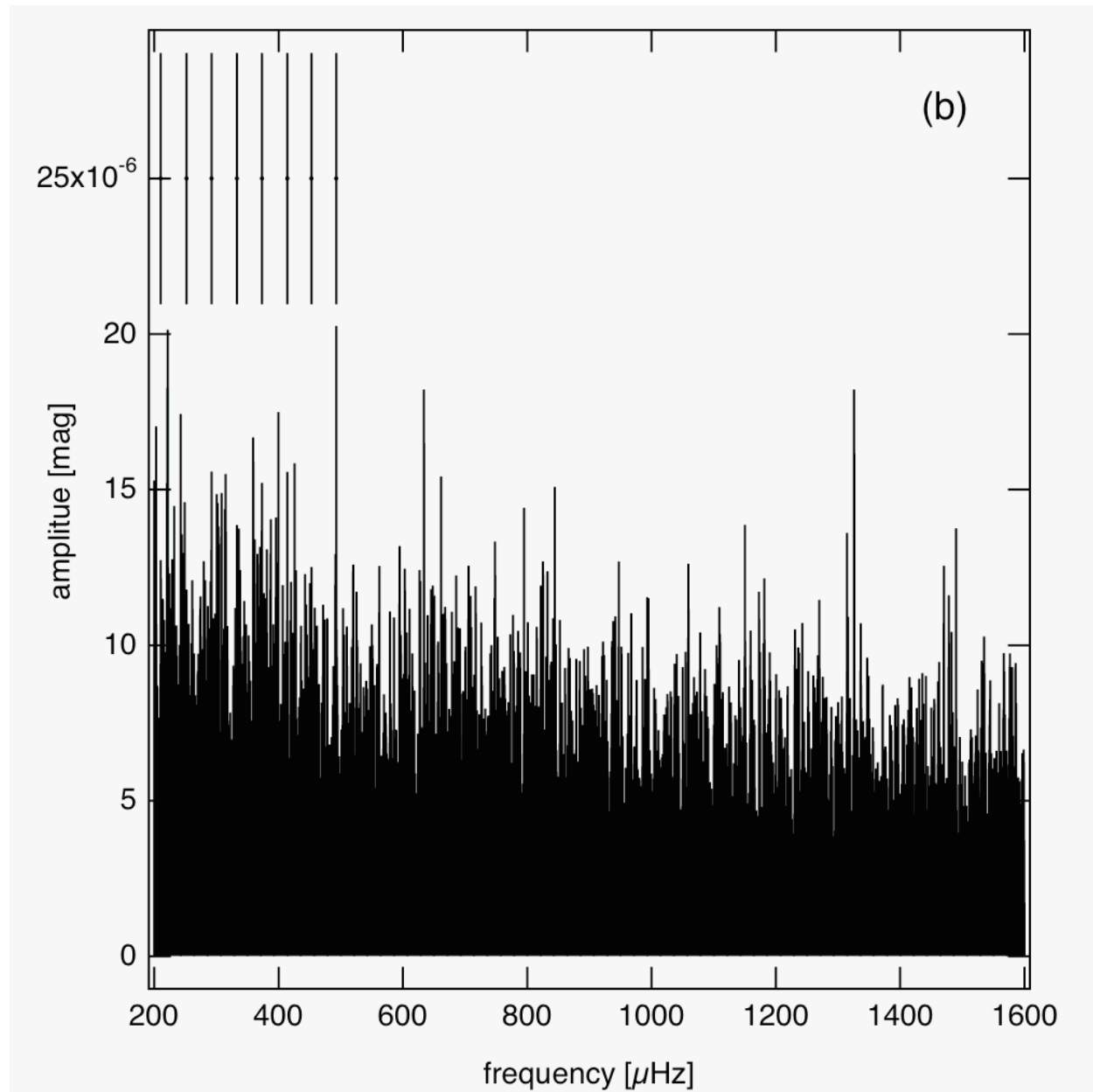

(c) Shows the significance spectrum peaks for significances greater than 4.0. Vertical lines indicate where the MOST 3–10 peaks are located.

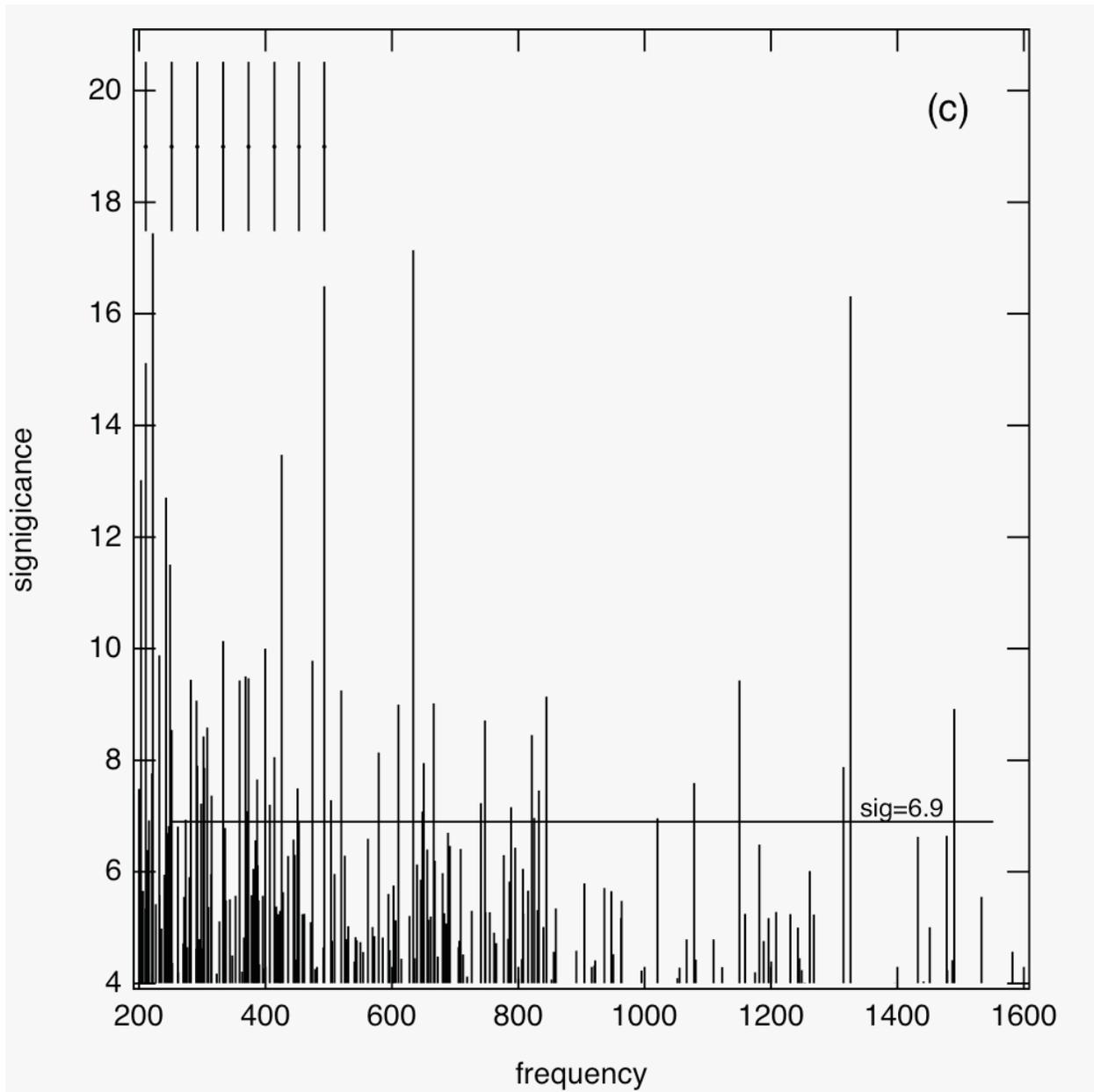

Fig. 2 — The raw spectrum peaks from MOST (with sig ≥ 6.9) are plotted in an echelle diagram with a folding frequency of 40 μHz. Labeled and encircled data points designate the MOST peaks that we believe could be $l = 0$ p-modes. The 1/1d alias corrected $l = 0$ p-modes from K2003 and C2005 are also shown.

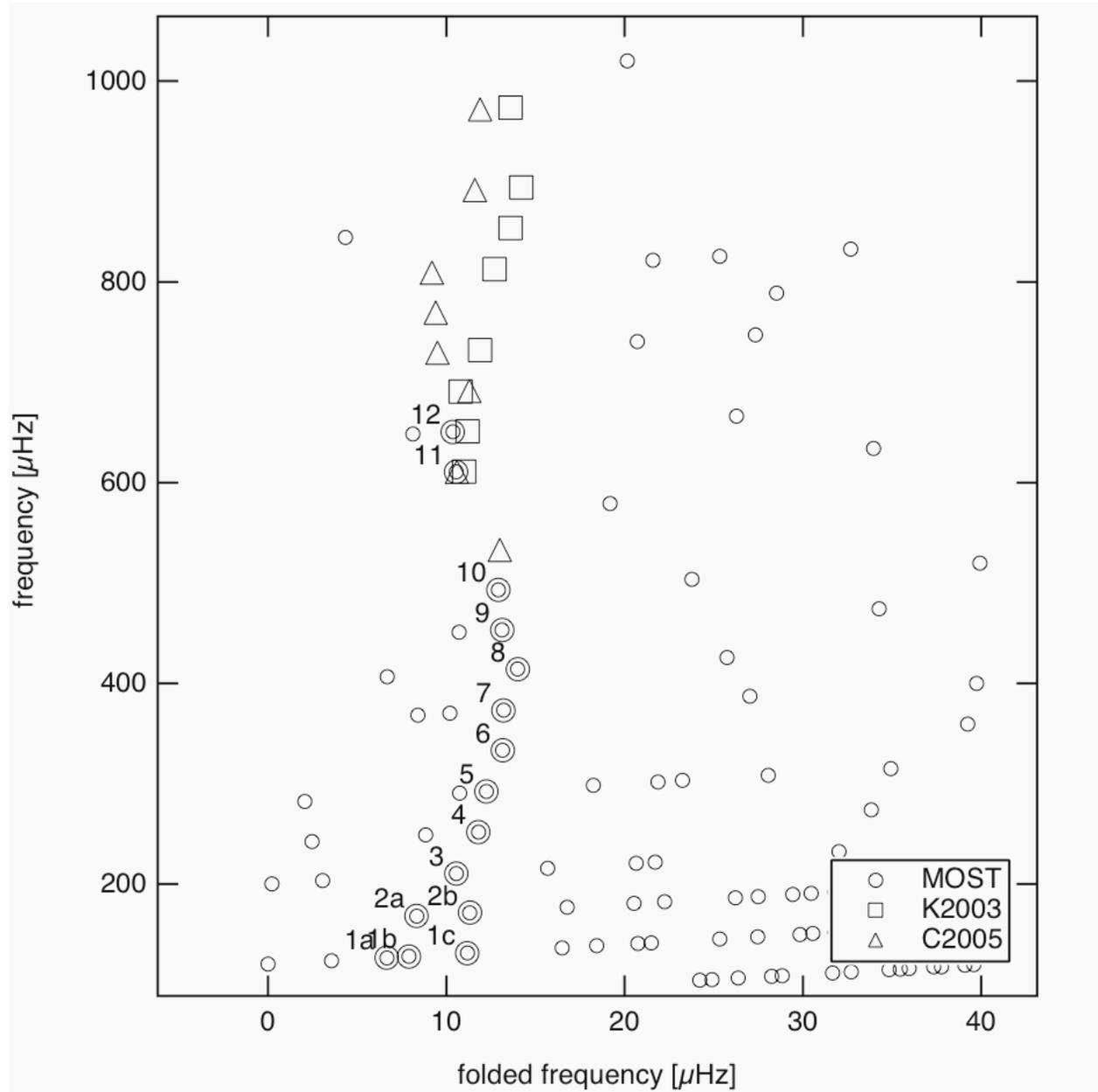

Fig. 3 — The uncorrected modes (small plot symbols) from K2003 and C2005 are plotted in an echelle diagram with folding frequency of 40 μHz. Large symbols show the location of the 1/1d alias corrected $l = 0$ p-modes. The scale of the plot is identical to Fig. 2.

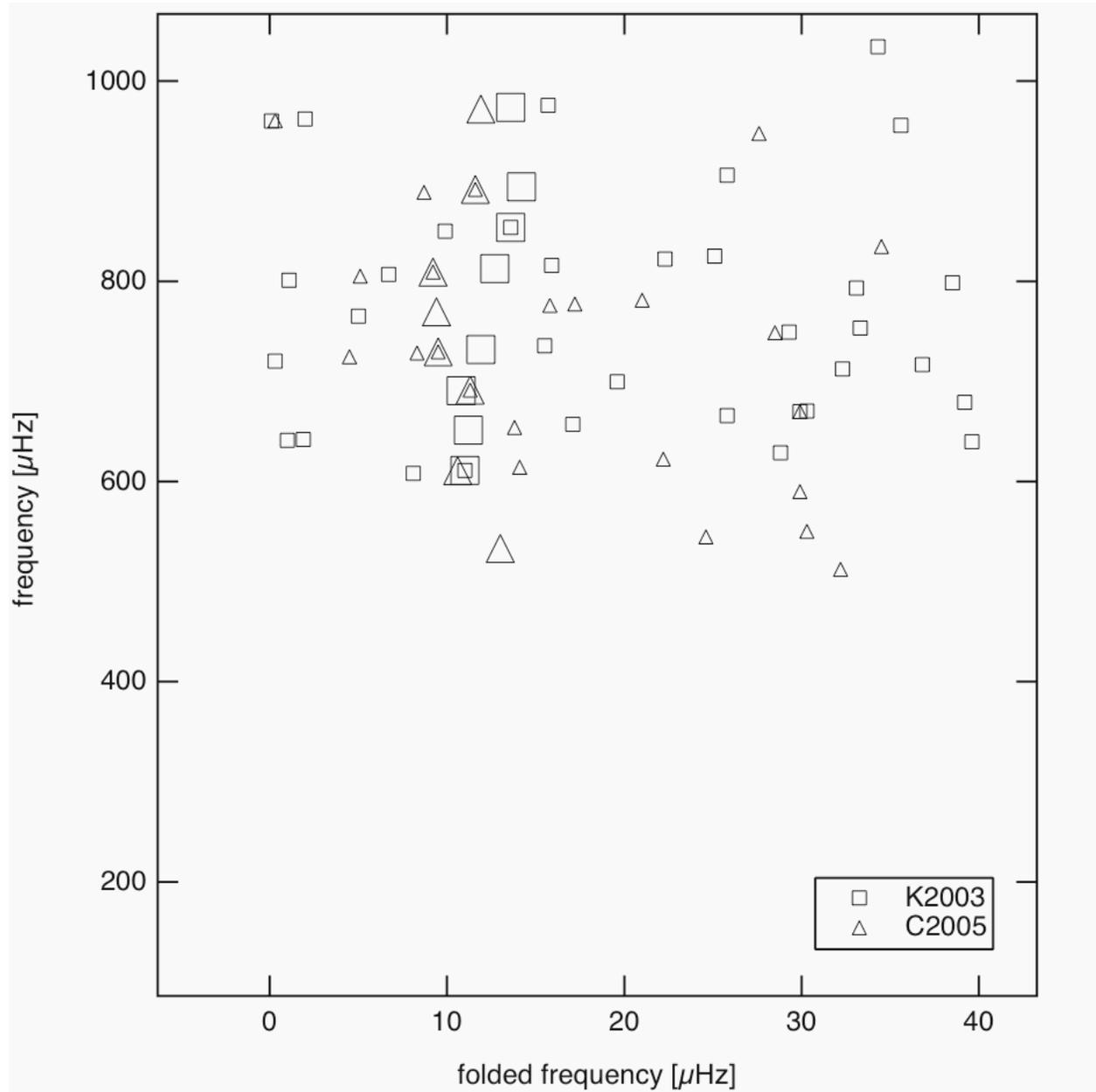

Fig. 4 — $\chi^2$ values ≤ 4.0 for models from the grid with $Z = 0.04$ and $X = 0.71$ are plotted as a function of mass, in $M_\odot$, and age, in Gyr. The MOST oscillation spectrum (3-10 modes) constrains the models to a curved two-dimensional sheet with a well defined minimum.

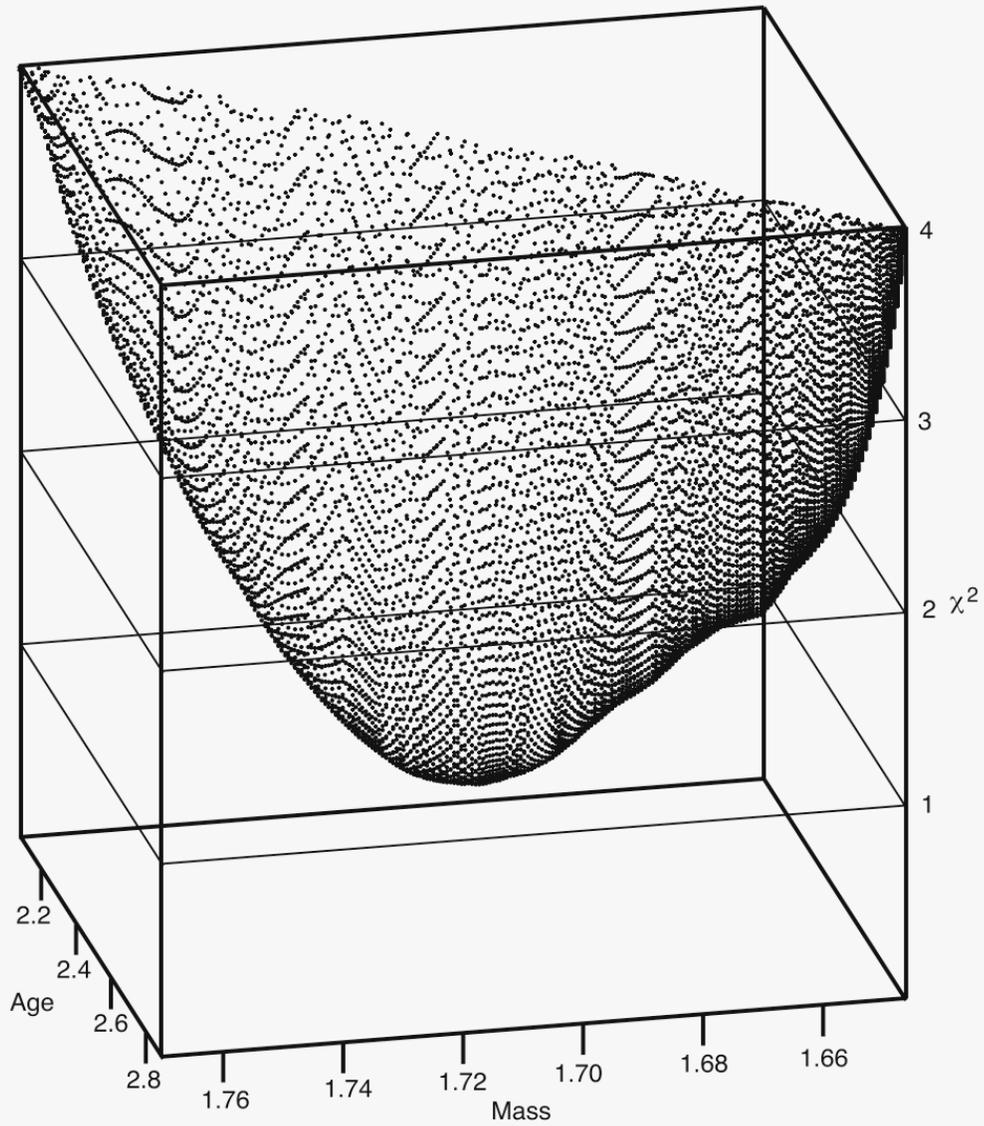

Fig. 5 — $\chi^2$ curves (see description in text) as a function of mass for the 3-10 combination of MOST spectrum peaks are shown for different combinations of metal and hydrogen mass fraction. Models with $\chi^2 = \sim 1$ correspond to models whose mode frequencies match the 8 MOST spectrum peaks to $\sim 0.1\%$.

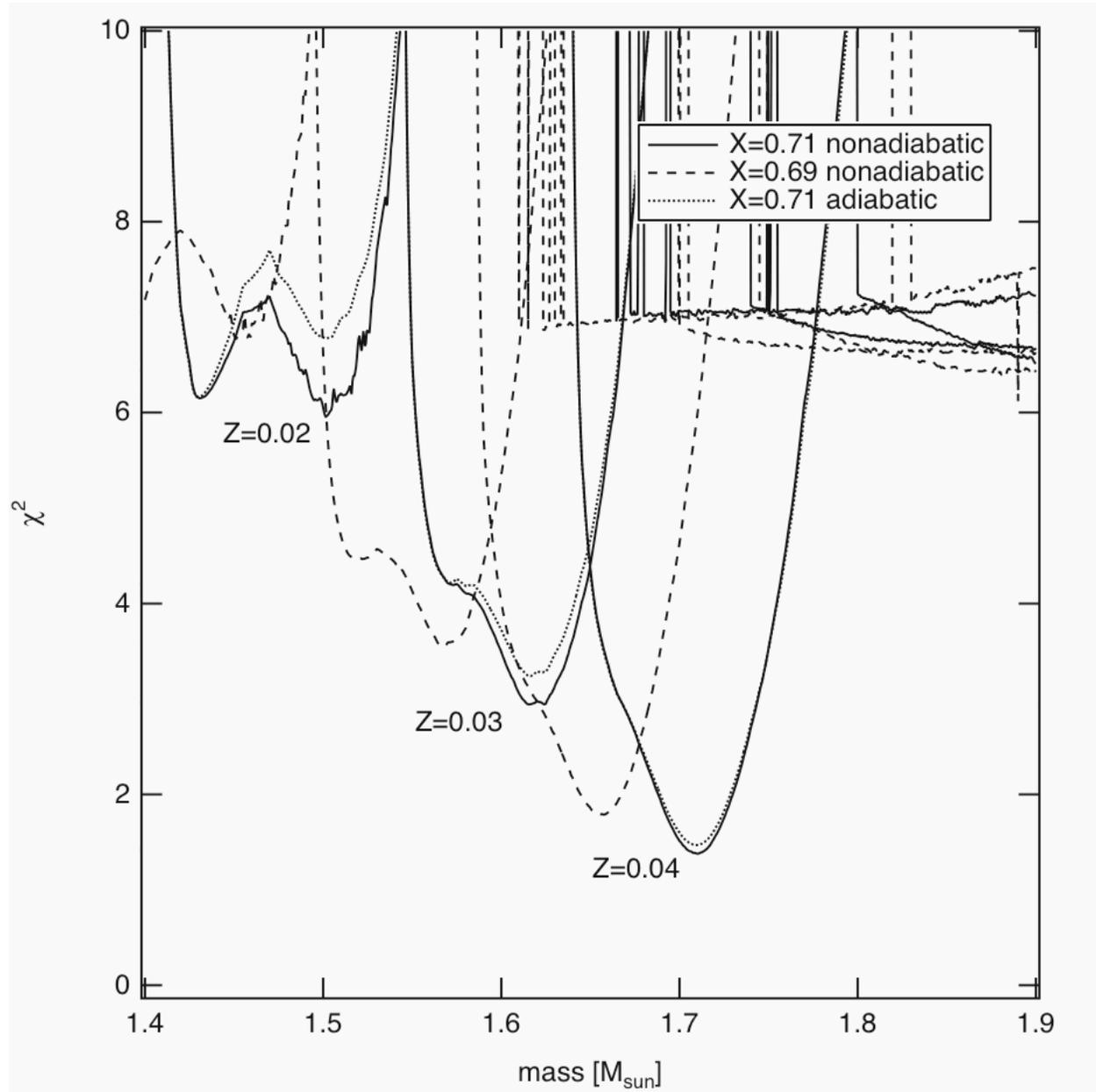

Fig. 6 — For each composition and for combinations of MOST and K2003 peaks, the surface temperature and luminosity of the model whose oscillation spectrum *best* matches the MOST spectrum is plotted in an HR-diagram. Two stellar evolutionary tracks are also plotted to suggest the evolutionary phase of the models. Models whose oscillation spectra best fit the 3-10 combination of MOST peaks are labeled with "zxx" indicating the metal, $Z = z/100$, and hydrogen, $X = xx/100$, mass fraction of the models. The two diagonal sequences of small data points show the range of models in the HR-diagram that fit the K2003 $l = 0$ oscillation data with $\chi^2 \leq 1$.

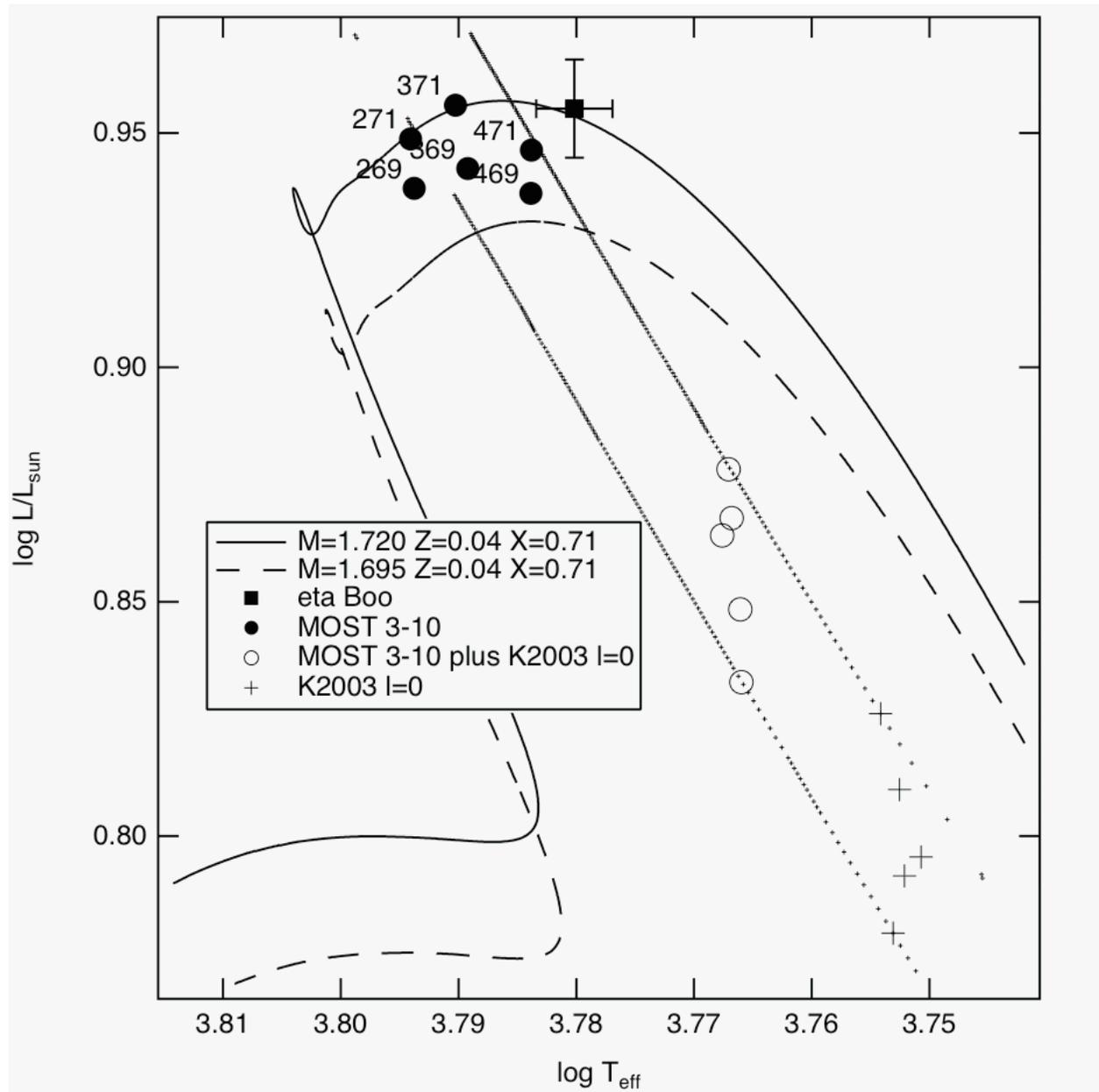

Fig. 7 — An echelle diagram with folding frequency of 40 µHz contains the position of the MOST spectrum peaks along with the $l = 0$, 1, and 2 $p$-modes of the model that best fits the MOST 3-10 combination of spectrum peaks. The adiabatic and nonadiabatic $l = 0$ $p$-mode frequencies are shown; only the nonadiabatic frequencies for the nonradial modes are plotted. The diagram also shows the $l = 1$ and $l = 2$ $g$-modes of the model that fall within the frequency range plotted.

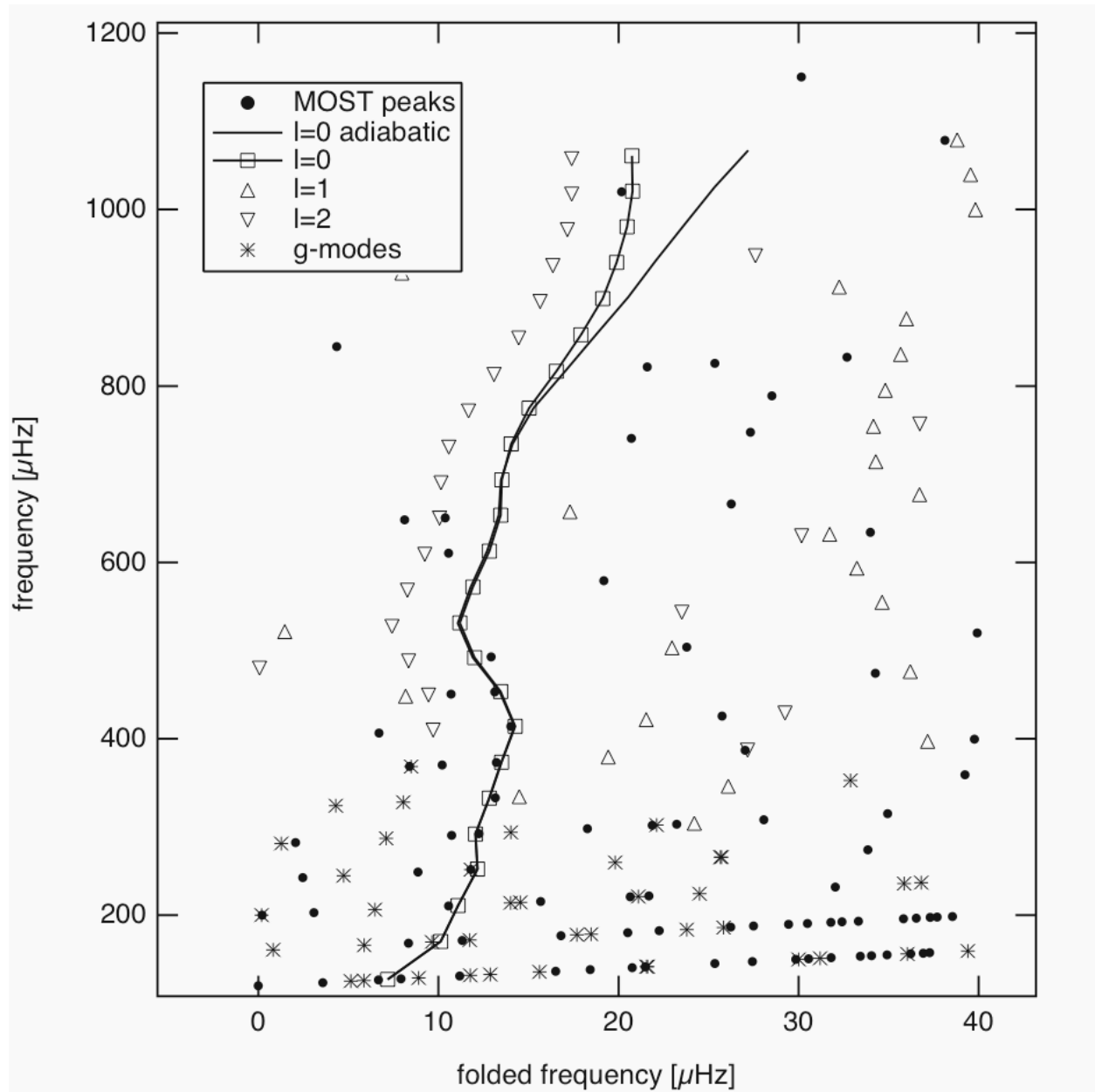

Fig. 8 — $\chi^2$ curves are plotted in projection as a function of mass for models from the grid with $Z = 0.04$ and $X = 0.71$. The different $\chi^2$ curves correspond to different combinations of MOST peaks as indicated in the legend and identified in Table 1.

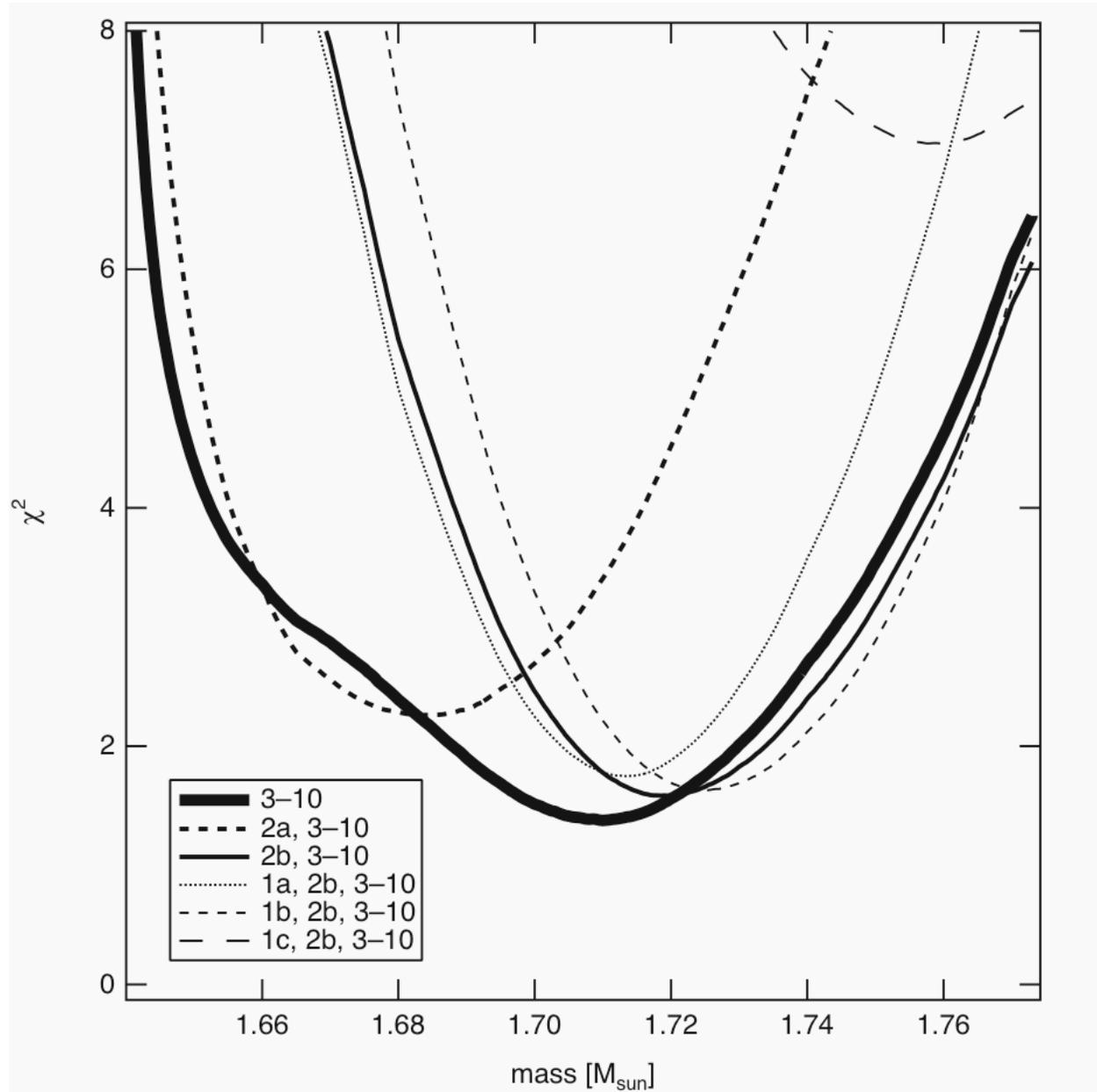

Fig 9 — The MOST and Kjeldsen $l = 0$ spectrum peaks are plotted in an echelle diagram with a folding frequency of 40 $\mu$Hz along with the adiabatic and nonadiabatic frequencies of the models that best fit the observed oscillation frequencies.

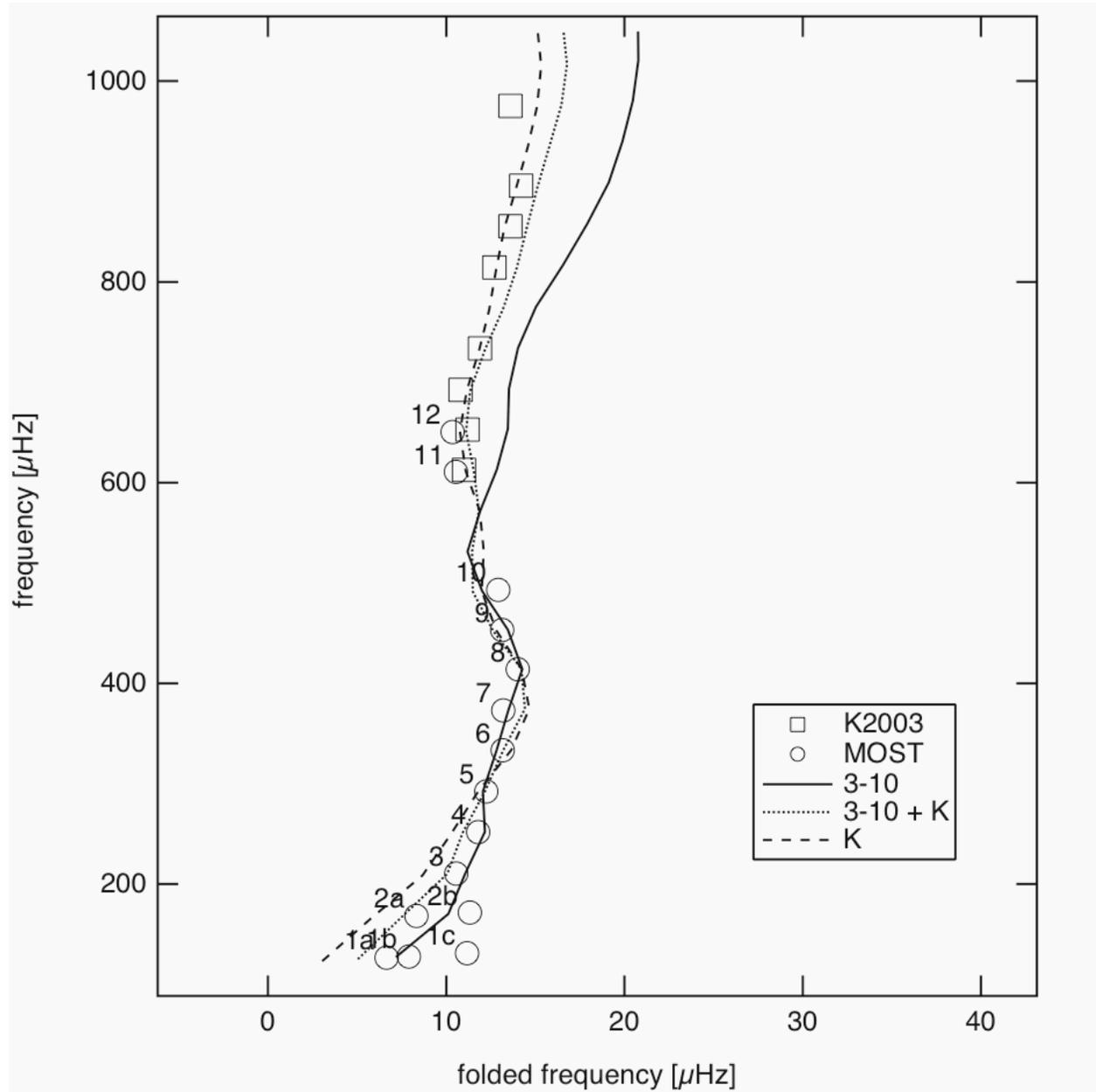

Fig. 10 — The $l = 0$ and $l = 1$ p-mode frequencies of a 1.710 $M_\odot$, $Z = 0.04$, $X = 0.71$ stellar model are plotted as a function of age. The frequencies of the 1b, 2b, 3-10 combination of MOST spectrum peaks are also plotted at 2.40259 Gyr, the age of the "3-10" model with the minimum $\chi^2$ (see Table 2). The fit to the $l = 0$ p-modes is very good. Note that below 300 $\mu$Hz, the discontinuous sequence of $l = 1$ modes are, in fact, g-modes (see text).

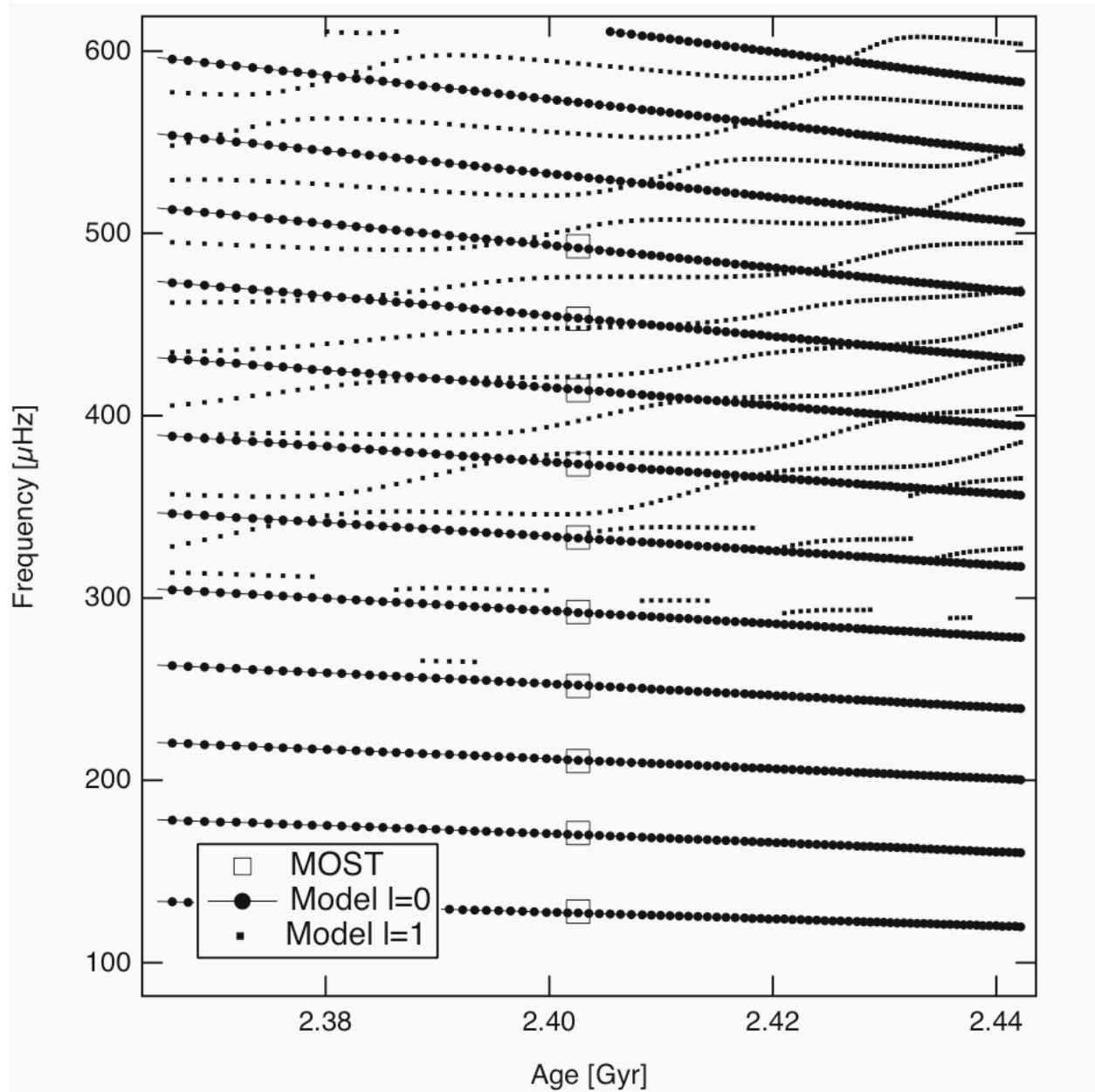

Fig. 11 — The three panels from top to bottom show in an echelle diagram with folding frequency equal to 40 $\mu$Hz, the $l = 0$, 1 and 2 $p$-mode frequencies for two nearly identical models, one with mass $M = 1.710$ M$_\odot$ and the other with $M = 1.715$ M$_\odot$. The background data points correspond to MOST peaks for η Boo with significances $\geq 6.9$.

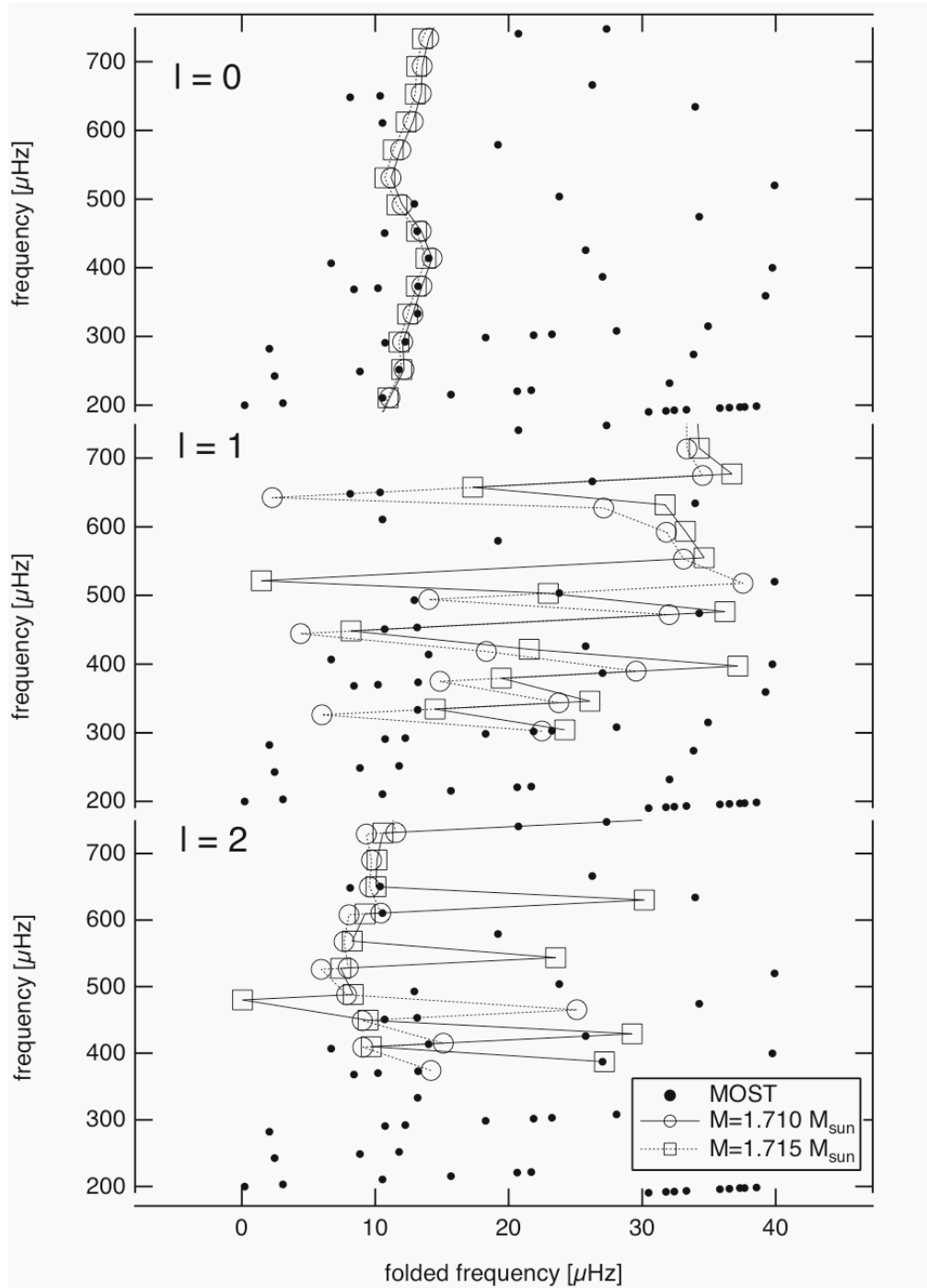

Fig. 12 — The number of model $l = 0$, 1, and 2 $p$-modes and $l = 1$ and 2 $g$-modes are counted in 80 $\mu$Hz wide bins and plotted in a histogram as a function of frequency. Similarly the number of MOST spectrum peaks with sig $\geq 6.9$ are also plotted in a histogram.

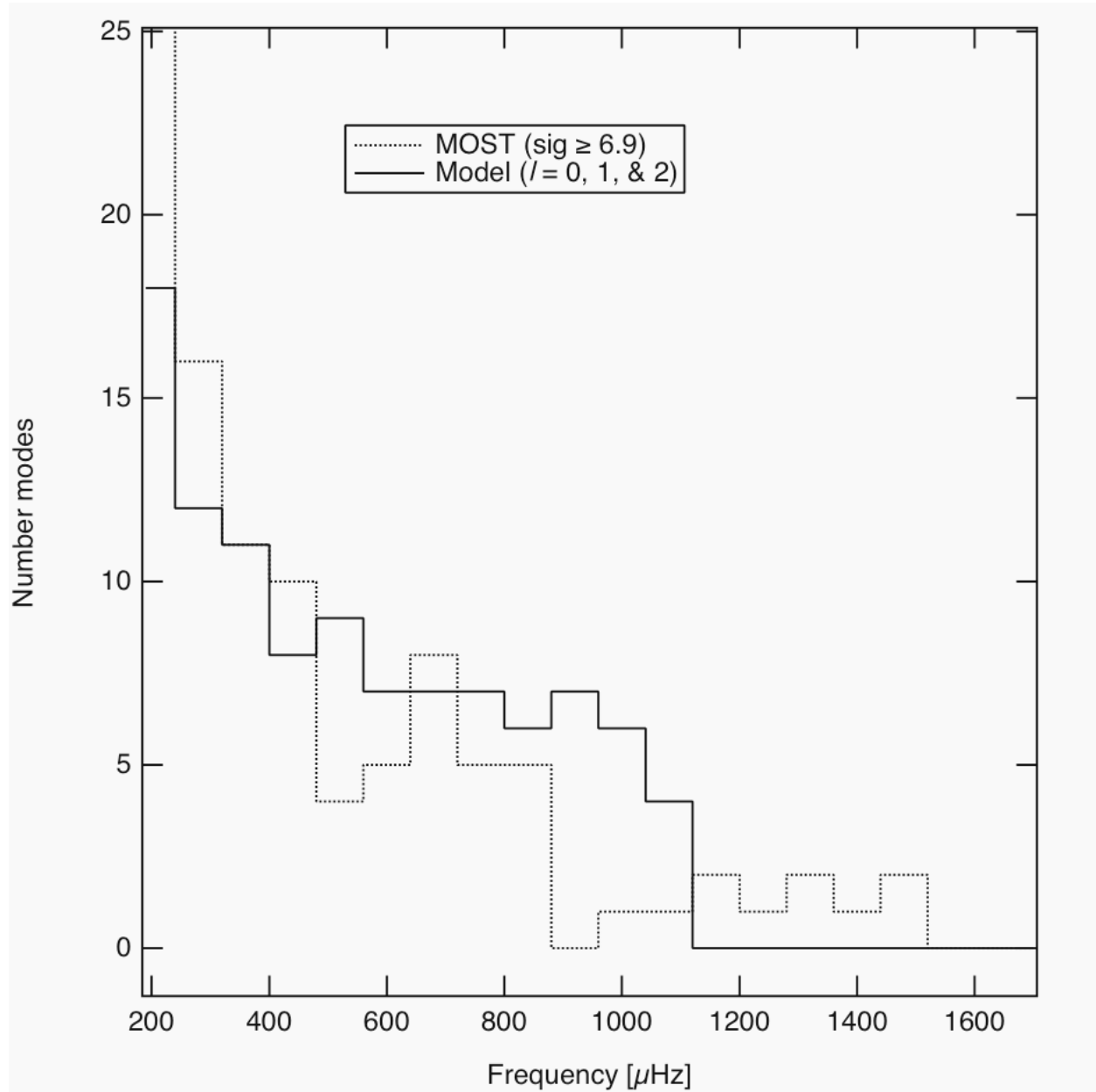

TABLE 1

| ID | Frequency [μHz] |
|---|---|
| 1a | 126.66 |
| 1b | 127.91 |
| 1c | 131.17 |
| 2a | 168.33 |
| 2b | 171.32 |
| 3 | 210.56 |
| 4 | 251.79 |
| 5 | 292.25 |
| 6 | 333.17 |
| 7 | 373.20 |
| 8 | 414.01 |
| 9 | 453.13 |
| 10 | 492.92 |
| 11 | 610.55 |
| 12 | 650.37 |

TABLE 2

| Peaks | X | Z | Mass [M$_\odot$] | Age [Gyr] | $\log T_{\text{eff}}$ | log L/L$_\odot$ | $\chi^2$ |
|---|---|---|---|---|---|---|---|
| 3-10 | 0.71 | 0.02 | 1.5015 | 2.41784 | 3.79407 | 0.94872 | 5.95420 |
| 3-10 | 0.71 | 0.03 | 1.6240 | 2.35747 | 3.79027 | 0.95591 | 2.94308 |
| 3-10 | 0.71 | 0.04 | 1.7100 | 2.40259 | 3.78381 | 0.94631 | 1.37696 |
| 3-10 | 0.69 | 0.02 | 1.4510 | 2.40704 | 3.79377 | 0.93822 | 6.77173 |
| 3-10 | 0.69 | 0.03 | 1.5680 | 2.34972 | 3.78920 | 0.94239 | 3.55573 |
| 3-10 | 0.69 | 0.04 | 1.6545 | 2.37742 | 3.78385 | 0.93715 | 1.79257 |
| 1b, 2b, 3-10 | 0.71 | 0.02 | 1.5015 | 2.41784 | 3.79407 | 0.94872 | 4.61737 |
| 1b, 2b, 3-10 | 0.71 | 0.03 | 1.6260 | 2.33880 | 3.79081 | 0.95804 | 2.35672 |
| 1b, 2b, 3-10 | 0.71 | 0.04 | 1.7255 | 2.32071 | 3.78774 | 0.96229 | 1.59583 |
| 1b, 2b, 3-10 | 0.69 | 0.02 | 1.4510 | 2.40704 | 3.79377 | 0.93822 | 5.25688 |
| 1b, 2b, 3-10 | 0.69 | 0.03 | 1.5745 | 2.31933 | 3.79126 | 0.95059 | 2.78947 |
| 1b, 2b, 3-10 | 0.69 | 0.04 | 1.6690 | 2.29564 | 3.78770 | 0.95282 | 1.75993 |
| 3-10 + K | 0.71 | 0.02 | 1.4290 | 2.98381 | 3.76599 | 0.83289 | 3.29027 |
| 3-10 + K | 0.71 | 0.03 | 1.5655 | 2.75501 | 3.76761 | 0.8642 | 2.57748 |
| 3-10 + K | 0.71 | 0.04 | 1.6610 | 2.71511 | 3.76708 | 0.87823 | 2.25961 |
| 3-10 + K | 0.69 | 0.03 | 1.5125 | 2.73143 | 3.76605 | 0.84836 | 2.62415 |
| 3-10 + K | 0.69 | 0.04 | 1.6085 | 2.67046 | 3.76679 | 0.86782 | 2.29031 |
| K | 0.02 | 0.71 | 1.4145 | 3.13735 | 3.75310 | 0.77933 | 0.12614 |
| K | 0.03 | 0.71 | 1.5470 | 2.92027 | 3.75073 | 0.79557 | 0.12813 |
| K | 0.04 | 0.71 | 1.6435 | 2.84108 | 3.75414 | 0.82610 | 0.12973 |
| K | 0.03 | 0.69 | 1.4975 | 2.87277 | 3.75212 | 0.79155 | 0.12605 |
| K | 0.04 | 0.69 | 1.5885 | 2.84391 | 3.75254 | 0.80996 | 0.13093 |